\documentclass[journal]{vgtc}                
\ifpdf
  \pdfoutput=1\relax                   
  \usepackage{graphicx}                
  \DeclareGraphicsExtensions{.pdf,.png,.jpg,.jpeg} 
\else
  \usepackage{graphicx}                
  \DeclareGraphicsExtensions{.eps}     
\fi%

\graphicspath{{figures/}{pictures/}{images/}{./}} 

\usepackage{microtype}                 
\PassOptionsToPackage{warn}{textcomp}  
\usepackage{textcomp}                  
\usepackage{mathptmx}                  
\usepackage{times}                     
\usepackage{cite}                      
\usepackage{tabu}                      
\usepackage{booktabs}                  

\onlineid{0}

\vgtccategory{Research}
\vgtcpapertype{application/design study}

\usepackage[final]{changes}

\usepackage{nicefrac,verbatim}
\usepackage{xfrac}

\usepackage{booktabs}
\newcommand{\ra}[1]{\renewcommand{\arraystretch}{#1}}
\usepackage{multirow}
\usepackage{caption,subcaption,graphicx}

\title{Interactive Visual Analysis of Structure-borne Noise Data}

\author{Rainer Splechtna, Denis Gra{\v{c}}anin, Goran Todorovi{\'{c}}, Stanislav Goja, Boris Bedi{\'{c}}, Helwig Hauser, and Kre{\v{s}}imir Matkovi{\'{c}}}
\authorfooter{
\item
Rainer Splechtna and Kre{\v{s}}imir Matkovi{\'{c}} are with the  VRVis Research Center.
E-mail: \{Splechtna$|$Matkovic\}@VRVis.at.
\item
Denis Gra{\v{c}}anin is with Virginia Tech in Blacksburg, VA.
E-mail: gracanin@vt.edu.
\item
Goran Todorovi{\'{c}}, Stanislav Goja, and Boris Bedi{\'{c}} are with AVL-AST d.o.o. Zagreb, Croatia.
E-mail: \{goran.todorovic$|$stanislay.goja$|$boris.bedic\}@avl.com.
\item
Helwig Hauser is with the University of Bergen, Norway. 
E-mail:~Helwig.Hauser@UiB.no.
}

\shortauthortitle{Splechtna \MakeLowercase{\textit{et al.}}: Interactive Visual Analysis of Structure-borne Noise Data}

\abstract{Numerical simulation has become omnipresent in the automotive domain, posing new challenges such as high-dimensional parameter spaces and large as well as incomplete and multi-faceted data.  
In this design study, we show how interactive visual exploration and analysis of high-dimensional, spectral data from noise simulation can facilitate design improvements in the context of conflicting criteria.  
Here, we focus on structure-borne noise, i.e., noise from vibrating mechanical parts.  
Detecting problematic noise sources early in the design and production process is essential for reducing a product's development costs and its time to market.  
In \added{a} close collaboration of visualization and automotive engineering, we designed a new, interactive approach to quickly identify and analyze critical noise sources, also contributing to an improved understanding of the analyzed system.  
Several carefully designed\replaced{, interactive linked}{ and linked, interactive} views enable the exploration of noises, vibrations, and harshness at multiple levels of detail, both in the frequency and spatial domain.  
This enables swift and smooth changes of perspective; selections in the frequency domain are immediately reflected in the spatial domain\added{,} and vice versa.  
Noise sources are quickly identified and shown in the context of their neighborhood, both in the frequency and spatial domain.  
We propose a novel drill-down view, especially tailored \replaced{to}{for} noise data analysis.
Split \replaced{boxplots}{box-plots} and synchronized 3D geometry views support comparison tasks\deleted{, as well}.
With this solution, engineers iterate over design optimizations much faster, while maintaining a good overview at each iteration.  
We evaluated the new approach in the automotive industry, studying noise simulation data for an internal combustion engine.  
} 


\CCScatlist{ 
 \CCScat{K.6.1}{Management of Computing and Information Systems}%
{Project and People Management}{Life Cycle};
 \CCScat{K.7.m}{The Computing Profession}{Miscellaneous}{Ethics}
}

\begin{document}

\firstsection{Introduction}

\maketitle

Simulation has become indispensable in automotive engineering.
Continuous improvements in computing and storage technology make it possible to simulate ever larger and more complex models.  
Still, compromises between a detailed representation and some degree of approximation are necessary.  
The engineers' experience and intuition remain of paramount importance when working with data from simulation.  
For an optimal cooperation, engineers need advanced exploration and analysis technology that supports their cognitively demanding tasks.  
\par 
Ever stricter noise regulations and requirements of increased passenger comfort are the main driving forces in noise, vibration, and harshness (NVH) studies.  
Two main sources of noise and vibrations are moving mechanical parts and air flow around an object.  
In this paper\added{,} we focus on \deleted{the} vibrations\added{,} caused by moving parts that can be excited externally (\added{e.g., }by the road\deleted{, e.g.}) or internally (\added{e.g., }by \added{the} engine or \added{the} transmission system\deleted{, e.g.}).
While \deleted{at lower frequencies these}\added{such} vibrations can \replaced{lead to}{produce} an unpleasant shaking of cabin parts \added{at lower frequencies}, \deleted{the} vibrations at higher frequencies produce so-called structure-borne noise.  
NVH simulation aims at capturing such situations, generating large amounts of challenging data (multi-dimensional, complex structure, etc.).
Due to time constraints and limited computational and storage resources, data is not generated for all frequencies and all engine operating points (at varying load, speed, etc.). 
The engineers' expertise is indispensable in the study of \added{such} sparse NVH simulation data~-- not only to ``fill in'' the gaps.  
Such an analysis usually leads to further design improvements, targeting the identified issues with noise, vibrations, and/or harshness, then followed by a new cycle of simulation and analysis.  
\par  
Here, we present~-- as the result of a long-term collaboration between automotive industry and visualization research~-- a novel approach to the exploration and analysis of structure-born noise.  
We used a participatory design approach to identify the requirements for an effective and efficient analysis workflow and abstracted the corresponding analysis tasks.  
Through iterative design we identified suitable choices for visualization and interaction.  
The new approach serves as a valuable complement to standard approaches, such as the Campbell diagram~\cite{Campbell-1922-a}
(see Fig.~\ref{fig:campbell_drill_down}, left), providing highly valuable opportunities for interactive drill-down.  
With the new approach, analyzing NVH data has become faster,  
and, at the same time, we also record an improved comprehension of the underlying phenomenon and better insight in physical aspects of the studied system.  
\par 
Noise simulation is situated in two related domains, frequenc\replaced{ies}{y} and \replaced{space}{spatial}, \replaced{both}{each} with particular, non-trivial hierarchical structure: (1)~The basic units in the frequency domain are individual harmonics.
Generally, \added{a} harmonic is a sinusoidal signal\added{,} whose frequency is an integer multiple of \replaced{a particular, }{the} fundamental frequency. In our case\added{,} the fundamental frequency \replaced{is}{represents} the given engine speed\added{,} expressed in Hz. The \added{intrinsically} complex engine vibrations \replaced{are given}{can be represented} as a superposition of \replaced{numerous}{infinite number of} sinusoidal (harmonic) vibrations. \replaced{In NVH simulation, we compute sufficiently many harmonics to successfully capture all relevant vibration effects.}{In practice we can calculate only a finite number of harmonics: this number must be large enough to capture vibration effects we want to analyze.} 
%
While our hearing range spans 16000\,Hz, only a representative selection of frequencies  between 30 and 4000\,Hz is \added{actually} simulated.  
\replaced{Since d}{D}ealing with hundreds of harmonics\deleted{, however,} is still infeasible\replaced{,}{ so} the data is \added{further} aggregated into frequency bands\replaced{:}{,} \nicefrac{1}{3} octaves\deleted{,} and octaves.  
(2)~In our case, the spatial domain is the surface geometry of an engine consisting of surface cells.
The cells are grouped into disjunct parts that are meaningful for the analysis.
All parts together represent the engine.
Respecting this semantically important hierarchical structure of both the frequency and the spatial domain imposes an interesting challenge when designing an appropriate exploration and analysis approach.  
\par 
To provide an informative overview of noise contributions of \added{the} different engine parts at different frequencies for different aggregation levels, and to enable an efficient interactive drill-down, e.g., to localize important NVH features, or to explore the spatial neighborhood of problematic faces, we arrived at a solution with \added{several carefully designed,} coordinated multiple views.  
While the individual views of our approach not really amount to a radical innovation in fundamental visualization research, their careful design, both in terms of visualization mapping and interaction design, \replaced{make}{do provide} an interesting case of successfully supporting the joint exploration and analysis of data that is given with respect to two \added{tightly related and} semantically sub-structured domains.
\replaced{Accordingly, o}{O}ur main contributions are:~ 
\textbf{1.}~\replaced{Our t}{T}ask analysis and abstraction for the \replaced{study}{analysis} of NVH simulation data with the potential to generalize, at least partially, to related cases with two \added{related} data domains \replaced{with}{of} non-trivial structure.~ 
\textbf{2.}~\replaced{Our new,}{An} integrated interactive visual analysis solution with carefully \replaced{designed}{chosen and customized} views\replaced{, providing}{ which provides} an efficient way to \replaced{work out}{solve} all identified tasks \replaced{in the study}{and requirements 
for the analysis} of NVH simulation data.
\par 
Very positive feedback from NVH experts, and the significant speedup of the analysis demonstrate the usefulness of our approach.
Partially abstracting from the very concrete case of NVH studies, we identify opportunities for generalizing selected aspects of our new approach to other, related scenarios, where data is given with respect to both spatial locations and frequencies, for example, in stress analysis, non-destructive testing, or axial waves propagation. 

\section{Structure-Borne Noise Analysis}

%

\begin{figure}[t!]
\centering
\includegraphics[width=\columnwidth]{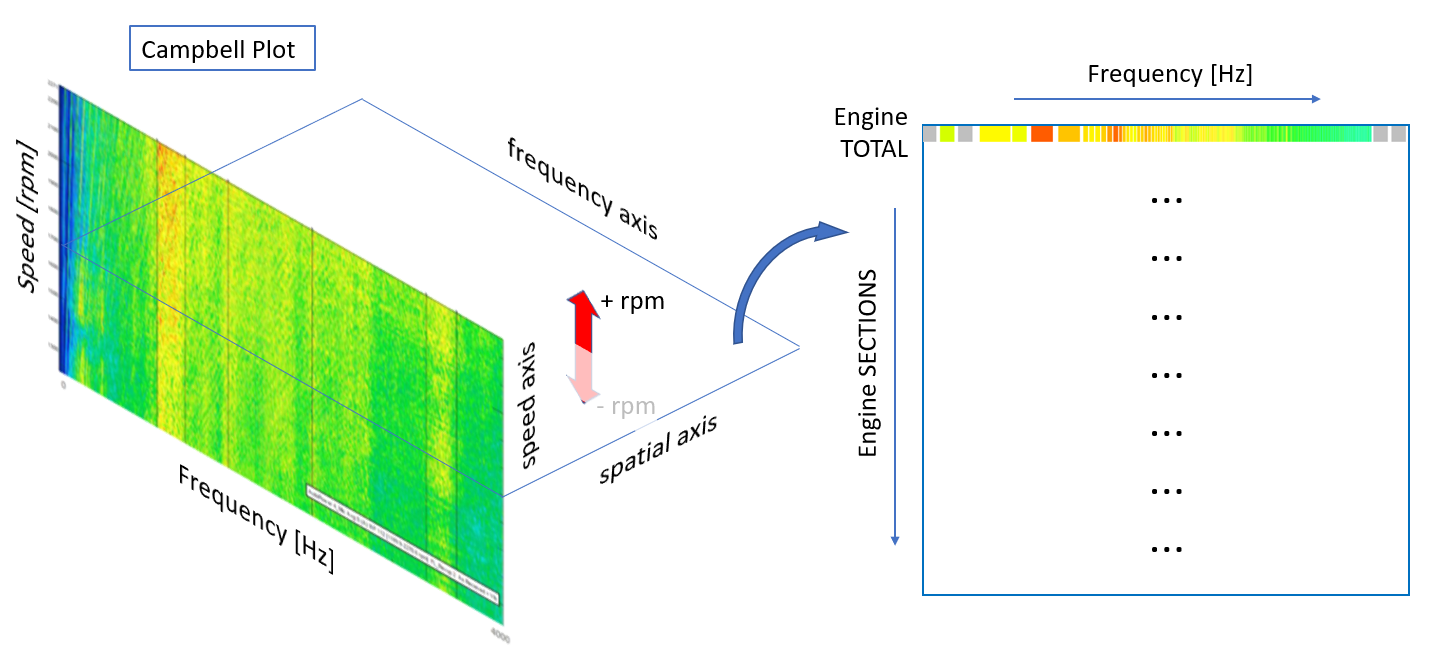}
\caption{The Campbell diagram on the left shows aggregated data for every frequency and engine speed\deleted{s}. Critical spots have to be examined \added{separately} using \replaced{all of the}{complete} data.}
\label{fig:campbell_drill_down}
\end{figure}

An important step in the automotive design process of a new product (e.g., a car or an engine) is \added{a} {N}oise, {V}ibration and {H}arshness analysis.  
In the automotive industry, NVH is an umbrella term that covers noise and vibration suppression, the suppression of squeaks, rattles, and `tizzes', as well as sound design~\cite{Harrison-2004-a}. 
Given an increased awareness of noise pollution, combined with an ongoing shift from internal combustion engines towards electric or hydrogen engines, NVH has gained increased importance as many parts have to be re-designed~\cite{Di-Marco-2019-a}.  
NVH analysis is performed for practically all automotive  components~\cite{Mandke-2019-a,Shaik-Mohammad-2017-a,Shaik-Mohammad-2019-a,Prajith-2020-a}. 
%
%
\par 
The goal of \replaced{an}{the} NVH analysis is twofold: first, the analysis should check that the product's exterior noise will be below certain thresholds to satisfy the given noise pollution legislation; second, the analysis should reveal whether interior noise\added{s} and vibration\added{s} \replaced{are}{is} minimized to increase \added{the} vehicle occupants' comfort and driving experience.
\replaced{It also}{Additionally, it} provides data for the expected durability of the product.
\par 
Challenges in analyzing, understanding, and solving NVH issues stem from the multi-dimensional character of the considered scenario and the interplay of the involved system components.  
Any NVH analysis~\cite{Souksavanh-2020-a} should be conducted early to  help designers with identifying potential NVH issues as soon as possible.
Usually, the analysis is repeated in different phases of the design.  
%
%
\par  
In the context of NVH, \textbf{Noise} describes audible sound, with particular attention \deleted{paid} to the frequency range from 30 to 4000\,Hz. 
Two main \deleted{contributing} sources of noise are: 
\textbf{structure-borne noise}\replaced{, due to}{ caused by} vibrating structural surfaces\replaced{, either caused}{ (induced by} internal\added{ly} (e.g., \added{by} engine vibrations) \replaced{or}{and} external\added{ly} \deleted{sources} (e.g., \added{due to} road surface roughness); and
\textbf{air-borne noise}\added{,} resulting from fluid pressure fluctuations\replaced{ that are}{,} transferred to \replaced{the}{a} vehicle's structure (e.g., \added{by} flow turbulence around an open roof window).
\par 
\textbf{Vibration} describes \added{a} tactile vibration in the frequency range of 30--200\,Hz~\cite{Wang-2010-a}. 
Vibrations at higher frequencies contribute mainly to noise; vibrations at lower frequencies can produce discomfort.  
\par 
\textbf{Harshness} describes \replaced{our}{human} perception of the quality and transient nature of noise and vibration.
While noise and vibration can be objectively measured\replaced{, being}{  (}physical quantities\deleted{)}, harshness is a 
subjective measure.
%
%
%
%
\par 
The main contributors to a vehicle's interior noise are powertrain noise, road noise, and wind noise.  Taking into account the increasing demands for lighter and more powerful vehicles, powertrain-induced noise plays a particularly important role~\cite{Wang-2010-a}.
%
%
%
\par 
A key part of a typical \deleted{workflow in} automotive NVH analysis is \textbf{structure-borne noise analysis}, targeting the higher end of the vibration spectrum.  
Data from a time-domain multi-body simulation is analyzed at several engine speeds, followed by a frequency response analysis with special focus on the vibration of the model's outer surface, responsible for noise radiation.
In order to quantify vibration, the normal-to-surface velocit\replaced{ies}{y} of the outer surface are computed.

\added{The \textbf{velocity level} $(L_v)$ compares the calculated normal-to-surface velocity ($v$) to a reference velocity value ($v_0$) on a logarithmic scale and is expressed in decibels [dB]:
\begin{equation} \label{eq:1}
L_v = 20 \cdot \log_{10}(\nicefrac{v}{v_0})
\end{equation}
The \textbf{discrete velocity level} is the velocity level at a point on the object’s surface. The \textbf{integral velocity level} is calculated for an entire surface by integrating the discrete velocities over the surface and applying Equ.~\ref{eq:1}.
}
\added{Velocity levels and noise levels are related in a complex way: the fluid(s) between the vibrating object and the point of interest (air, water, etc.), the vibrating object’s geometry, the distance from the vibrating object, and several other factors have to be considered. 
}

\added{Particular \textbf{integral velocity level limits} (in dB) serve as initial 
criteria (acceptable\,/\,borderline\,/\,unacceptable). These limits are derived from numerous noise measurements of engines of the same engine class (size, power, etc.). They also take \added{into account} \replaced{hearing}{the our ears} sensitivity to different frequencies \deleted{into account} and are therefore defined per frequency-band. \textbf{Discrete velocity level limits} are obtained from the integral velocity level limits by normalizing them to the unit surface area. This way, the factor of engine size is eliminated so that these discrete level limits are useful for evaluating velocity levels at different engine surface points.}

\deleted{The velocity level is 20 times logarithm of ratio between the normal-to-surface velocity and a reference velocity, expressed in decibels [dB].
Discrete velocity level is the velocity level calculated for a point at an object's surface. The integral velocity level is calculated for a surface and is obtained by integrating discrete velocities over the surface and recalculating them into velocity levels as described above.}

\deleted{Velocity level and noise levels are correlated in a complex way considering fluid properties (air, water, \ldots) between the vibrating object and the point in space of interest, vibrating object's geometry, distance from the vibrating object, etc. There exist numerical packages that take an object's vibrating surface velocities as input in order to calculate the sound intensity/levels at a spatial point of interest.}

\deleted{The frequency-dependent integral velocity level limits serve as design acceptance criteria. They are derived from numerous measurements of air-borne noise for an engine. They take into account human ear sensitivity to different frequencies and  additional factors. The discrete velocity level limits are obtained from the integral velocity level limits by removing factor of engine surface area, so they can be applied to evaluating velocity levels calculated at an engine surface spot.}

When some of the surface normal velocities are unacceptable, the design is changed and the analysis is repeated. Upon satisfying \added{the} prescribed surface velocity level\replaced{s}{ criteria} and in combination with the analyst's expertise, the design can be accepted by proving that the noise levels are within allowed ranges\added{,} either computationally or experimentally.
\par   
The simulation results in data with a particular structure, both with respect to frequencies, as well as geometrically.  
For a particular engine speed and for each cell on the outer surface, the velocity levels for different frequencies are computed.  
The number of cells ranges from tens or hundreds of thousands for common geometries to millions of cells for more intricate cases.  
The geometry is \deleted{semantically} subdivided either orientation-based (cells facing in a certain direction), or semantically (faces belonging to a certain engine part).  
The frequencies are sampled from all 4000 harmonics of the audible frequency range; altogether, several hundreds of harmonics are computed. 
These harmonic values are then aggregated into the frequency bands. 
Octave\deleted{ band}s are used for the highest level of aggregation, and \nicefrac{1}{3} octaves are used for a finer aggregation. Typically, there are about ten octaves, and three times as many \nicefrac{1}{3} octaves. 
%
In this paper, we use simulation results for an internal combustion engine operating at 2000\,RPM. The engine surface has more than 13\,000 vertices. Velocity levels are computed for each vertex for 359 harmonics and 26  \nicefrac{1}{3} octaves. More details on \added{the} data are provided in Sect.~\ref{sec:sim_data}.

\par 
To get \replaced{an}{a useful} overview in the frequency domain, \added{the} simulation results are typically aggregated into a Campbell diagram~\cite{Boyce-2012-a} (\deleted{e.g., }Fig.~\ref{fig:campbell_drill_down}\added{,} left).
This diagram depicts the entire model's integral velocity levels \deleted{distribution} across \replaced{all}{the} calculated engine speed\added{s} and \added{all} frequenc\replaced{ies}{y range}.  
For each engine operational point and for each frequency there is one color-coded location in the plot. All results that are computed over all cells for a single engine speed are shown \replaced{as}{with} one line. 
%
%
%
Information on potential ``hot spots'', as indicated by the Campbell diagram, serves as entry point for a more detailed analysis of their severity, causes\added{,} and possible means to resolve them, if proven critical.
%
\par 
Each such spot must be checked against frequency-dependent integral velocity level limits\replaced{ and}{,} then examined to understand the cause of the high surface velocities. Depending on the findings, the design has to be changed \deleted{in order} to reduce the velocity levels.
\added{In order to examine the behavior of the system in detail, the highly aggregated overview information provided in the Campbell diagram has to be unfolded in the frequency domain as well as in the spatial domain.}
This is a tedious and time-consuming process, which usually requires several tools for analysis and visualization.
\deleted{On one hand side, the highly aggregated Campbell diagram information has to be unfolded in frequency domain as well as in the spatial domains.} 
Computed velocity levels across frequencies are usually examined using 2D charts and statistics tools\replaced{. For the  analysis of the velocity levels in the spatial domain, the values are mapped on the engine model itself using a 3D visualization tool.}{, followed by 3D visualization of velocity levels on the engine model
itself} Linking the frequenc\replaced{ies}{y} and the spatial domain is usually left to engineers, who switch between tools to work out suggestions for a design improvement.  
\par 
\replaced{To accelerate and improve this tedious process, w}{W}e propose a new, integrated interactive visual analysis approach, enabling an efficient in-depth exploration of NVH data that helps \deleted{engineers} with identifying problematic locations and frequencies.  
Additionally, our approach contributes to improving the expert's insight into an engine's NVH behavior.  
\section{Related Work} 
\label{sec:related}
Numerical simulations usually generate large,  multi-dimensional data.\deleted{ sets.}
Exploration of such data \deleted{sets} is often difficult and require\added{s} innovative visualization approaches~\cite{Bonneau-2003-a}. 
Advanced frameworks and solutions are needed to support \added{the} interactive specification and visualization\deleted{s} of relevant features in the data~\cite{Bonneau-2003-a}. 
\par 
Usually, the interactive visual analysis of simulation data uses customized views that provide insight into all relevant facets of  multi-dimensional and spatio-temporal data~\cite{KH2012}.
However, not all visualization techniques, tools and views are well-suited for the study of simulation data.
\replaced{The s}{s}election of \added{appropriate} visualizations\replaced{, given a particular}{ based on different} simulation \deleted{application} domain\deleted{s} and \added{according} data characteristics\added{,} can be informed by a taxonomy of visualization solutions.
\added{Vernon-Bido et al.~\cite{Vernon-Bido-2015-a} describe such a taxonomy in the context of modeling and simulation, discussing issues of each case, and serving as a basis for the identification of tasks and requirements in our case. }
\added{Visualization in the automotive domain includes computer-aided-design, virtual reality, and scientific visualization~\cite{Stevens-2007-a}. 
The experts' domain knowledge can be integrated into a visual analytics system by machine learning and interactive labeling~\cite{Eirich-2020-a,Langer-2021-a}.
Further examples of visual analytics in the automotive domain include the Cardiogram~\cite{Sedlmair-2011-a} and an electrical distribution systems~\cite{Dominguez-2019-a}. As we deal with acoustic data, it was not feasible to apply solutions for other simulation types without adaptation. }
\par 
An expressive analysis of NVH data must respect the physical  representation of the simulated system and the used modeling and simulation methods.
For example, acoustic data from acceleration tests are used to conduct a detailed analysis of engine sub-components and determine the primary symptom of certain  errors~\cite{Eirich-2021-a}. 
\added{Although this work deals with acoustic data, the main analysis goal and data structure is different from our scenario. Therefore, the proposed solutions could not be directly applied to our case. }
Computer aided engineering (CAE) is central to the simulation of noise and vibration~\cite{Hampl-2010-a,Prajith-2020-a}.
\added{The lack of an integrated, interactive analysis tool for NVH data was one key motivation for our new approach.} 
\deleted{The developed visualization solutions must be carefully chosen to efficiently support the identified tasks.}
\deleted{Vernon-Bido et al.~\cite{Vernon-Bido-2015-a} describe a basic taxonomy of visualization solutions in the context of modeling and simulation and discuss issues of each case.
This taxonomy informed our identification of the tasks and requirements.}

\replaced{We needed a visualization solution that provides insight into hierarchical data in the frequency and the spatial domain, relying on the well proven concept of coordinated multiple views~\cite{CMV_STAR}. In addition to 2D views, we also needed 3D visualization. Our novel drill-down view relates to the Campbell diagram~\cite{Campbell-1922-a}, showing different engine parts against the different operating points.  Due to the hierarchical nature of our data, our solution also reflects Shneiderman's information visualization mantra~\cite{Shneiderman-1996-a}, making it possible to examine the data at various levels of detail. }{Quantitative Visualization requires specialized and customize views to provide insight into spatial and temporal/spectral features. We provide a novel drill-down view that includes customized Campbell diagram (frequency/speed)~\cite{Campbell-1922-a} and boxplots that supports data exploration  and Schneiderman's information visualization mantra~\cite{Shneiderman-1996-a}.}

\replaced{We realized several improvements of otherwise well-known representations to meet our needs. We adapted two-tone pseudo coloring~\cite{Saito-2005-a} to include several shades per hue and use it in a heat-map like display. We also enrich boxplots~\cite{tukey1977exploratory} with a background showing frequency-specific thresholds and we provide different sorting strategies when showing values at the finest level of details. Such adaptions follow from our needs and are as such not found in visualization for simulation yet. }{Seek and Find Visualization provides a clear representation of the input/output system of the simulation. For each view we provide a control panel to select the input/output parameters and support filtering data set for ranges of values. }

\replaced{We also provide animated 3D views of the model and use color to show (multiple) scalar fields and deformation to show radiation efficiency. }{Pattern and Flow visualization provides the dynamic display of evolving simulation. We provide an animated 3D view of an engine and use color to show (multiple) scalar fields.
The animated view uses deformation~\cite{Zheng-2002-a} to indicate the radiation efficiency. }

\replaced{Scalar field visualization methods, such as pseudocoloring, glyphs, and deformation, are also tightly related. Neeman et al.~\cite{Neeman-2005-a} discuss different scalar measures, thresholds, etc., to visualize tensors\,/\,multiple scalar fields from stress tensor fields in geomechanics. They provide a stress glyph that shows trends in a volume. Another approach to volume rendering using multi-dimensional transfer functions is discussed by Kniss et al.~\cite{Kniss-2001-a}. They identified a class of three-dimensional transfer functions for scalar data and provided the corresponding manipulation widgets. since our tasks do not require manipulation, for example, we limit our visualization to color coding the values, but we do provide various color scales, including a non-liner scale to stress out critical parts.
Finally, Zheng and Pang provide two visualization methods based on volume deformation~\cite{Zheng-2002-a} to show a continuous representation of tensor fields. Selected visualization tools also provide exaggerated deformation as a mean to visualize geometric variation. We deploy such an approach to communicate deformations of the 3D model. }{Scalar field visualizations include tensor visualization methods such as pseudocoloring, tensor glyphs, and deformation. Neeman et al.~\cite{Neeman-2005-a} discuss different scalar measures, thresholds, etc. to visualize tensor / multiple scalar fields applied to stress tensor fields from geomechanics. They provide a stress glyph that shows trends in a volume. Another approach to volume rendering using multi-dimensional transfer functions is discussed by Kniss et al.~\cite{Kniss-2001-a}. They identified a class of three-dimensional transfer functions for scalar data and provided the corresponding manipulation widgets. Zheng and Pang provide two visualization methods based on volume deformation~\cite{Zheng-2002-a} to show a continuous representation of tensor fields. } 

\deleted{Visualization in the automotive domain also includes computer-aided-design, virtual reality, and scientific visualization~\cite{Stevens-2007-a}. 
The experts' domain knowledge can be integrated into a visual analytics system by machine learning and interactive labeling~\cite{Eirich-2020-a,Langer-2021-a}.
Further examples of visual analytics in the automotive domain include the Cardiogram~\cite{Sedlmair-2011-a} and an electrical distribution systems~\cite{Dominguez-2019-a}.
}

\begin{table*}[t]
\caption{Overview of the results of our task and requirements analysis}
\centering
\ra{1.25} 
\begin{tabular}{@{}l@{~~~~}p{7.8cm}@{~~~~~}p{7.5cm}@{}p{0.1mm}@{}}
\toprule
\textbf{Levels}&\textbf{Tasks}&\textbf{Requirements}\\
\midrule
\textbf{Overview} 
& \raggedright\textbf{T1}: gain an overview of whether surface normal velocity levels are acceptable over \added{the} frequency bands as prescribed by external target levels and identify too noisy engine parts
& \raggedright\textbf{R1}: show fulfillment of externally prescribed target levels for frequency bands and engine parts at one glance (preattentively) 
&\\ 
\textbf{Medium} 
& \raggedright\textbf{T2}: identify prioritized parts to be examined for a possibly necessary improvement
& \raggedright\textbf{R2}: show distributions of velocity levels across frequency bands and engine parts, as well as information on the engine parts' size\added{s}, while maintaining the overview
&\\ 
\textbf{Detail}
& \raggedright\textbf{T3}: explore problematic frequency bands in context of their neighbours, both in the frequency and spatial domain, and explore problematic spatial locations and their neighbourhood, also in both domains, to pin-point parts to \replaced{improve}{fix}
& \raggedright\textbf{R3}: for any selected region in the frequency domain, \mbox{show the} corresponding cells in the spatial domain, \mbox{and for any} selected region of the engine, \mbox{show data in the frequency domain}
&\\ 
& \raggedright\textbf{T3a}:~examine all harmonics of a selected frequency band in the frequency domain;\phantom{x} 
  \textbf{T3b}:~examine most critical harmonics of a selected frequency band in the spatial domain;\phantom{x} 
  \textbf{T3c}:~explore radiation efficiency in the spatial domain;\phantom{x} 
  \textbf{T3d}:~identify acoustic hot-spots on the engine's surface
& \raggedright\textbf{R3a}:~show all harmonics in a frequency band in detail; support value-based comparisons, keep spatial coherence;\phantom{x} 
  \textbf{R3b}:~visualize the model geometry and show different velocity levels using color scales tailored to the task; \mbox{also show} model deformations 
&\\[1ex] \bottomrule 
\end{tabular}
\label{table:tasks}
\end{table*}

\section{Tasks and Requirements Analysis}
\label{sec:tasks}
\replaced{Our}{A} close collaboration of automotive engineers and visualization researchers, lasting \replaced{over}{for} one \added{and a half} year\deleted{and a half}, resulted in the following task abstraction, describing central parts of the engineers' workflow and therefore serving as the primary basis for our requirements analysis.  
The analysis process was carried out in numerous iterations, as well as the \replaced{corresponding}{according} design process.   
%
\par  
At the highest level (``Overview'' level in Table~\ref{table:tasks}),  engineers need a preattentive overview of whether the \added{relevant} \nicefrac{1}{3} octaves are acceptable with respect to the externally prescribed target levels; more specifically, they need to see whether the simulation results are acceptable, borderline, or unacceptable for each frequency band.  
They also need to identify critical engine regions, across the frequency domain, contributing the most to problems due to unacceptable velocity level limits (task T1).  This leads to \deleted{the following} requirement \deleted{(}R1\deleted{)}: 
for each frequency band and engine part show whether the externally prescribed target levels are fulfilled; this should be done in a preattentive form so that engineers can spot it instantly, i.e., without searching\deleted{ the visualization}.  
At this level, it becomes clear if the engine is too loud. 
\par 
Once problematic locations are identified, engineers need to dive deeper into the related physics to better understand the situation. 
At this medium level, the resolution of the velocity level scale increases.  
Three categories are not sufficient for this task, more details are needed while keeping the overview intact.  
Additional information is used to prioritize parts for a more detailed examination, and to contribute to an improved understanding of the simulated scenario (task T2).  \replaced{Here}{To support T2}, we need to show the distribution of velocity levels across frequency bands and engine parts, as well as \replaced{the size of the}{information on} engine parts\deleted{ size}, while still keeping the overview intact, i.e., showing all information at once so that critical locations can be prioritized (requirement R2).  
\par 
Thirdly, at the most detailed level, engineers have to explore problematic frequency bands and engine parts and all related harmonics in the frequency domain, and parts of the engine in the vicinity of the problematic cells in the spatial domain (task T3). 
To do so, we link both domains so that a selection in one domain can easily \replaced{lead to an examination}{be examined} in the other (requirement R3).
We show\added{,} for a selected problematic region\added{, the related spatial cells} in the frequency domain\deleted{ the related spatial cells}, and\added{,} for a selected region of the engine, the corresponding data in the frequency domain.  
\par  
The detailed exploration at this level consists of several sub-tasks. 
In the frequency domain, individual harmonics have to be compared to \deleted{the} other harmonics from the same frequency band (task T3a). To \replaced{do so}{support this task}, we have to show all harmonics from a frequency band in full detail, provide statistics for a selected harmonic, and examine how it relates to other harmonics, i.e., which others are similar and which are not (requirement R3a).  
Simultaneously, in the spatial domain, engineers need to get an overview of local noise levels checking \added{the} prescribed limits across several frequencies or frequency bands (task T3b), they have to identify so-called acoustic hot-spots on the engine's outer surface (task T3c), and explore \added{their} radiation efficiency, which is not explicitly computed, but can be estimated by experienced engineers based on \added{the} involved geometry parts and their deformations (task T3d). Tasks T3b--T3d lead to requirement R3b, i.e., that the model geometry needs to be shown \replaced{as well as}{and} different velocity levels using \replaced{carefully}{specially} tailored color scales, as well as model deformations.   
\par 
Table~\ref{table:tasks} summarizes our findings \replaced{of our}{during the} task and requirements analysis. 
\replaced{In the following, we}{We now} explain the design choices \added{that} we arrived \added{at} \replaced{in}{during} our iterative design process in order to meet the identified requirements.   
\section{Visual Encoding and Interaction Design}
\label{sec:approach}
In multiple\added{,} collaborative design iterations, we arrived at appropriate visualization and interaction solutions that satisfy the requirements as imposed by the identified \deleted{data} analysis tasks (Section~\ref{sec:tasks}).  
The so-called drill-down view acts as the central hub of the analysis and we motivate and describe its visual encoding first. 
\begin{figure}[t!]
    \centering
\begin{subfigure} {1\columnwidth}    
    \includegraphics[width=\columnwidth]{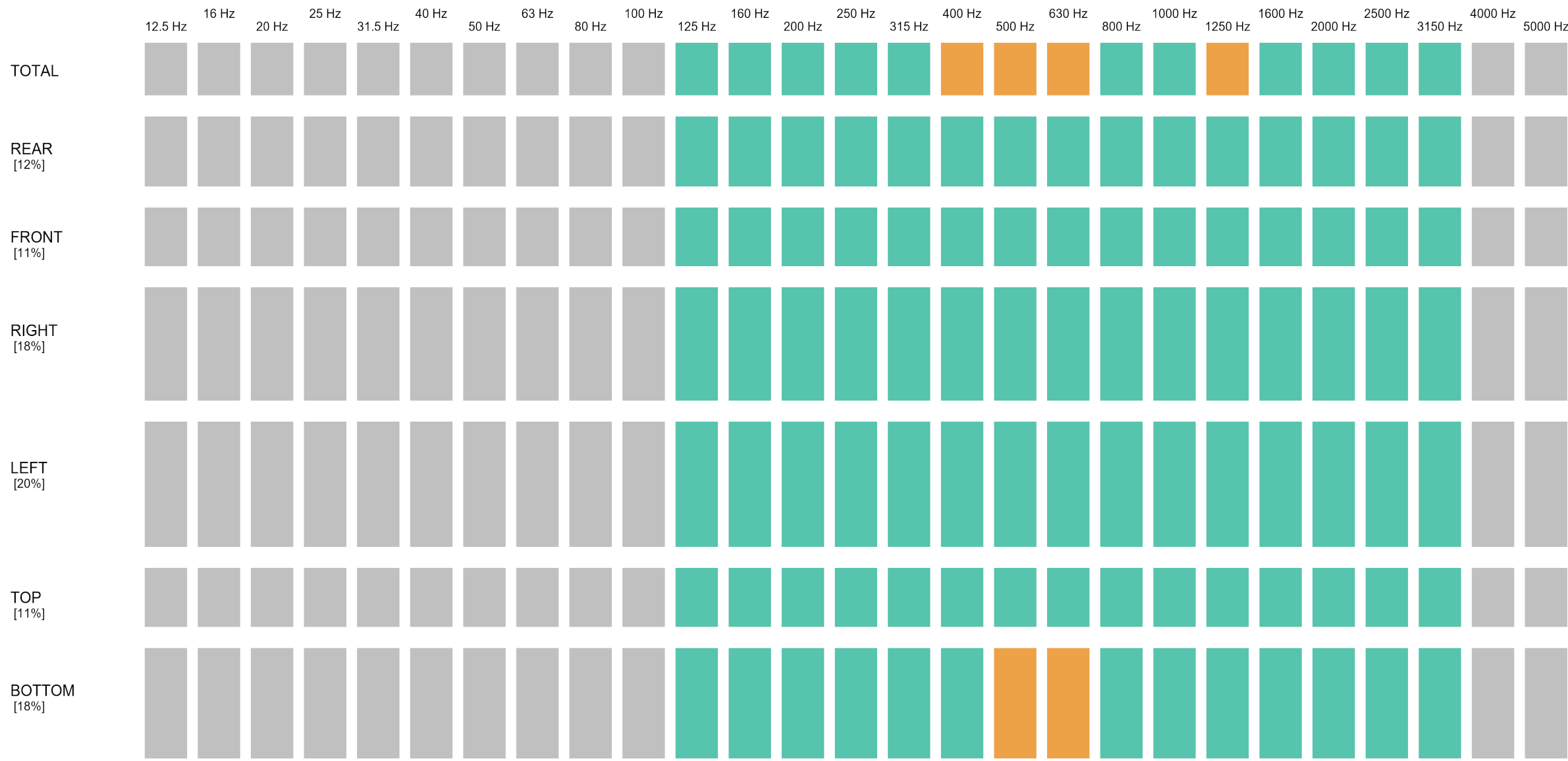}
    \caption{Overview-level analysis of integral velocity level acceptance according to externally prescribed levels. The yellowish grid cells indicate violations of the thresholds. Grid cells, i.e., \nicefrac{1}{3} octaves, without specified limits are grayed out.}
    \label{fig:overview_limits}
    \end{subfigure} \hfill
\begin{subfigure} {.48\columnwidth}
    \centering
    \includegraphics[width=\columnwidth]{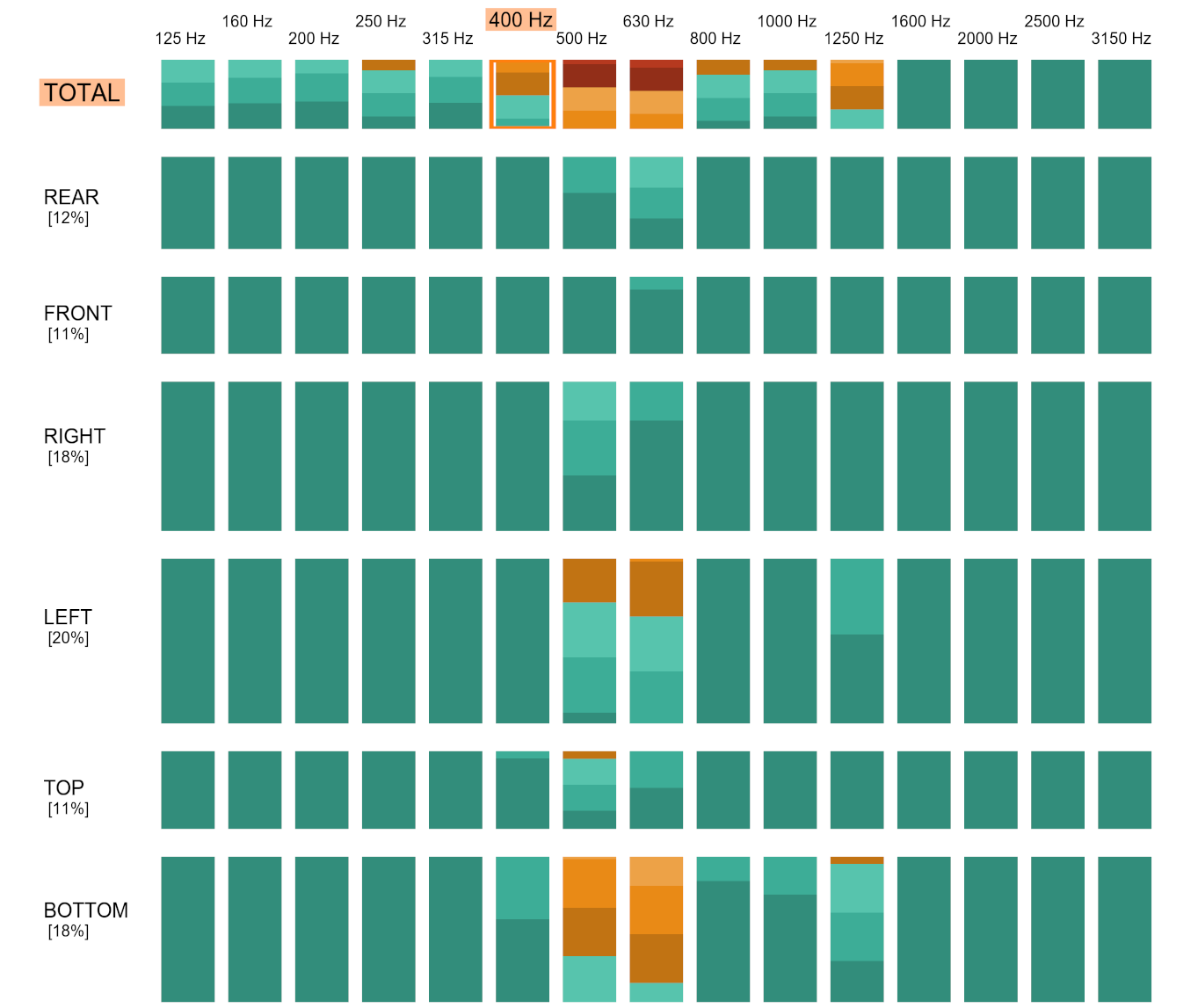}
    \caption{Medium-level analysis of integral velocity level acceptance. The same data as in Figure~\ref{fig:overview_limits} is visualized here. The single scalar integral velocity level values of the grid cells are shown using \added{our} two-tone pseudo coloring to better gauge the severity of the acceptance criterion violation.}
    \label{fig:overview_limits_horizon}
\end{subfigure} \hfill
\begin{subfigure} {.48\columnwidth}
    \centering
    \includegraphics[width=\columnwidth]{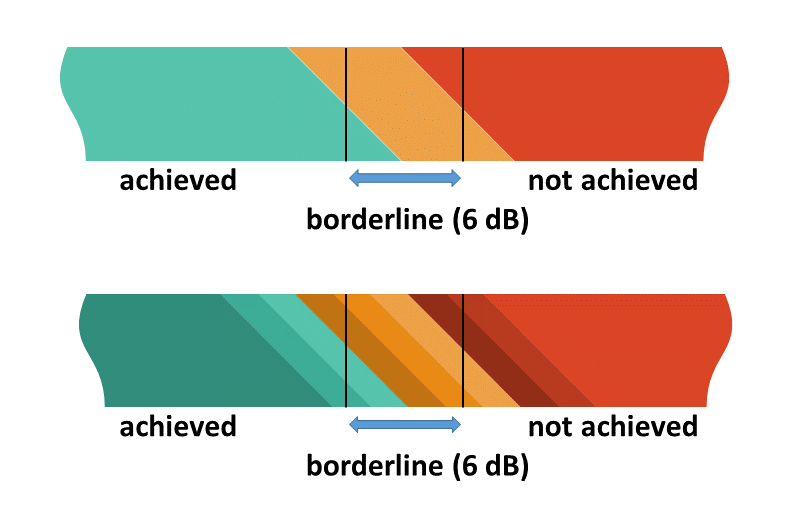}
    \vspace{.75mm}
    \caption{\added{Our} two-tone pseudo coloring encoding is applied relative to the individual borderline range of a given frequency band. The encoding can use \deleted{just} the traffic light tones (top) or multiple shades per traffic light tone, \replaced{here}{in this case} three shades (bottom).}
    \label{fig:limits_encoding}

\end{subfigure} \hfill

\begin{subfigure} {.48\columnwidth}
    \centering
    \includegraphics[width=1.\columnwidth]{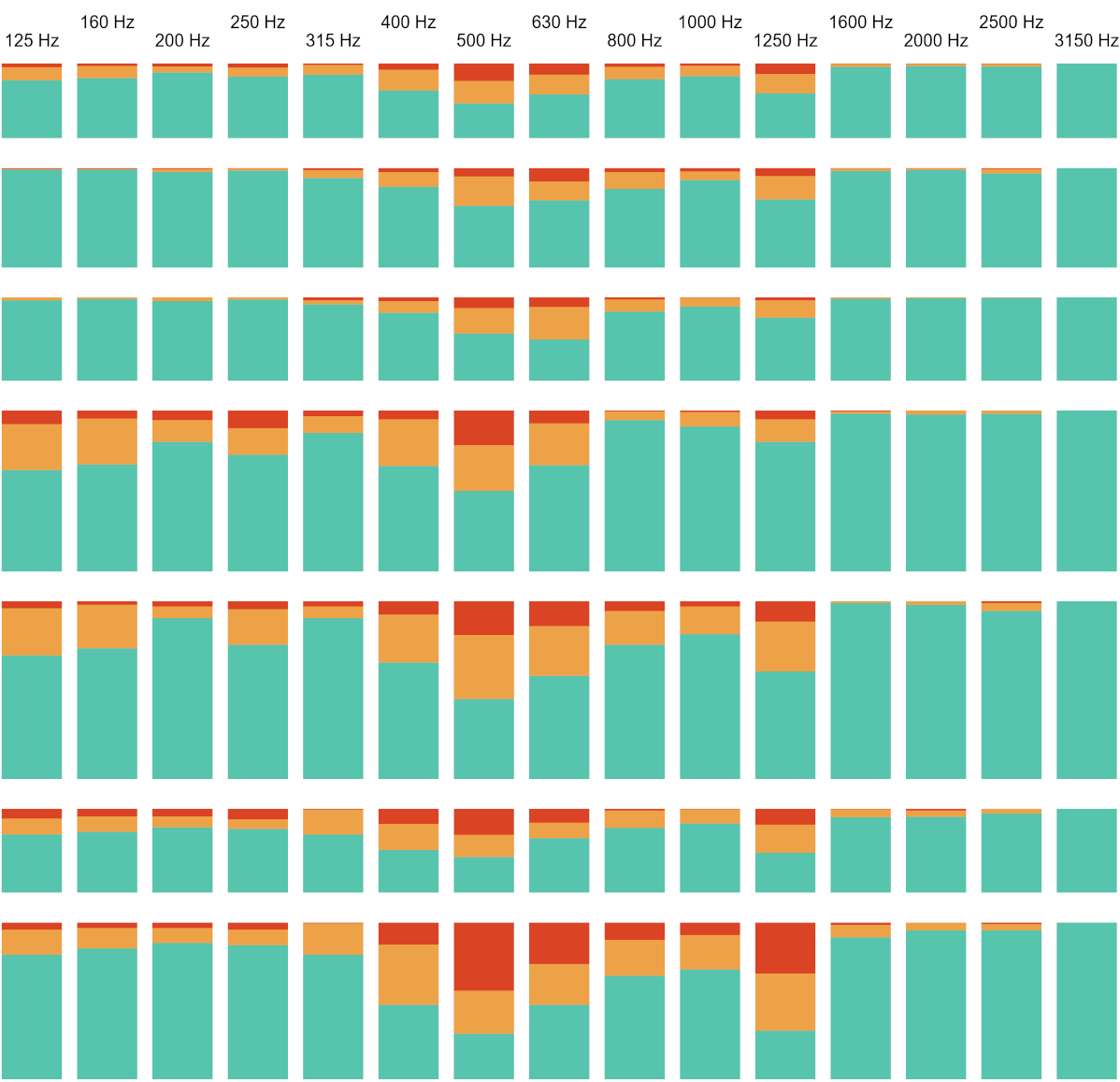}
    \caption{Medium-level analysis using area weighted and ranked discrete velocity levels using discrete velocity level acceptance thresholds for coloring.}
    \label{fig:overview_discrete_limits}
\end{subfigure} \hfill
\begin{subfigure} {.48\columnwidth}
    \centering
    \includegraphics[width=1.\columnwidth]{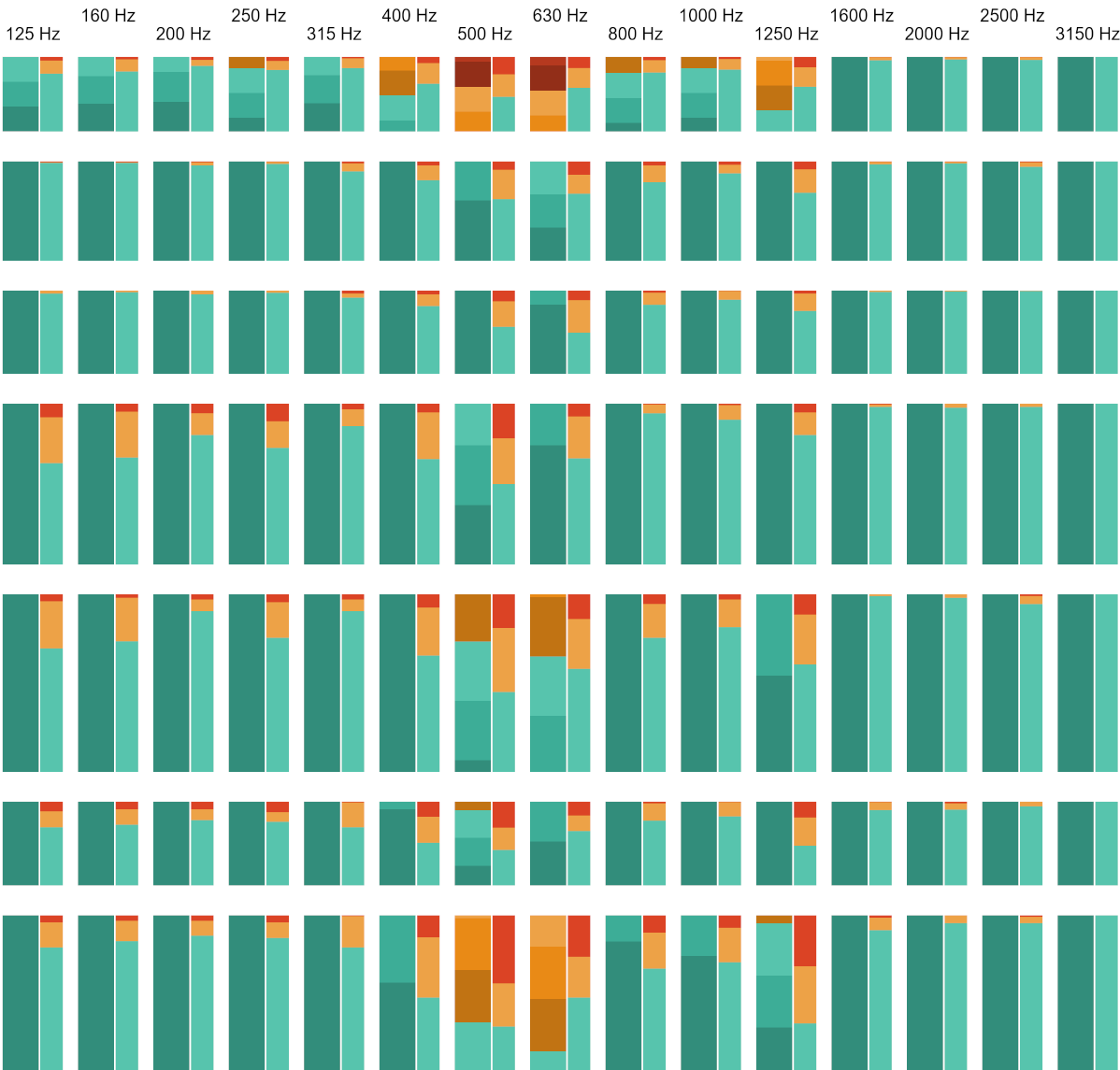}
    \caption{Medium-level analysis using combined velocity level acceptance, i.e., showing Fig.~\ref{fig:overview_limits_horizon} and Fig.~\ref{fig:overview_discrete_limits} side by side in one grid cell.}
    \label{fig:overview_combined}
\end{subfigure} \hfill

\begin{subfigure} {\columnwidth}
    \centering
\includegraphics[width=\textwidth]{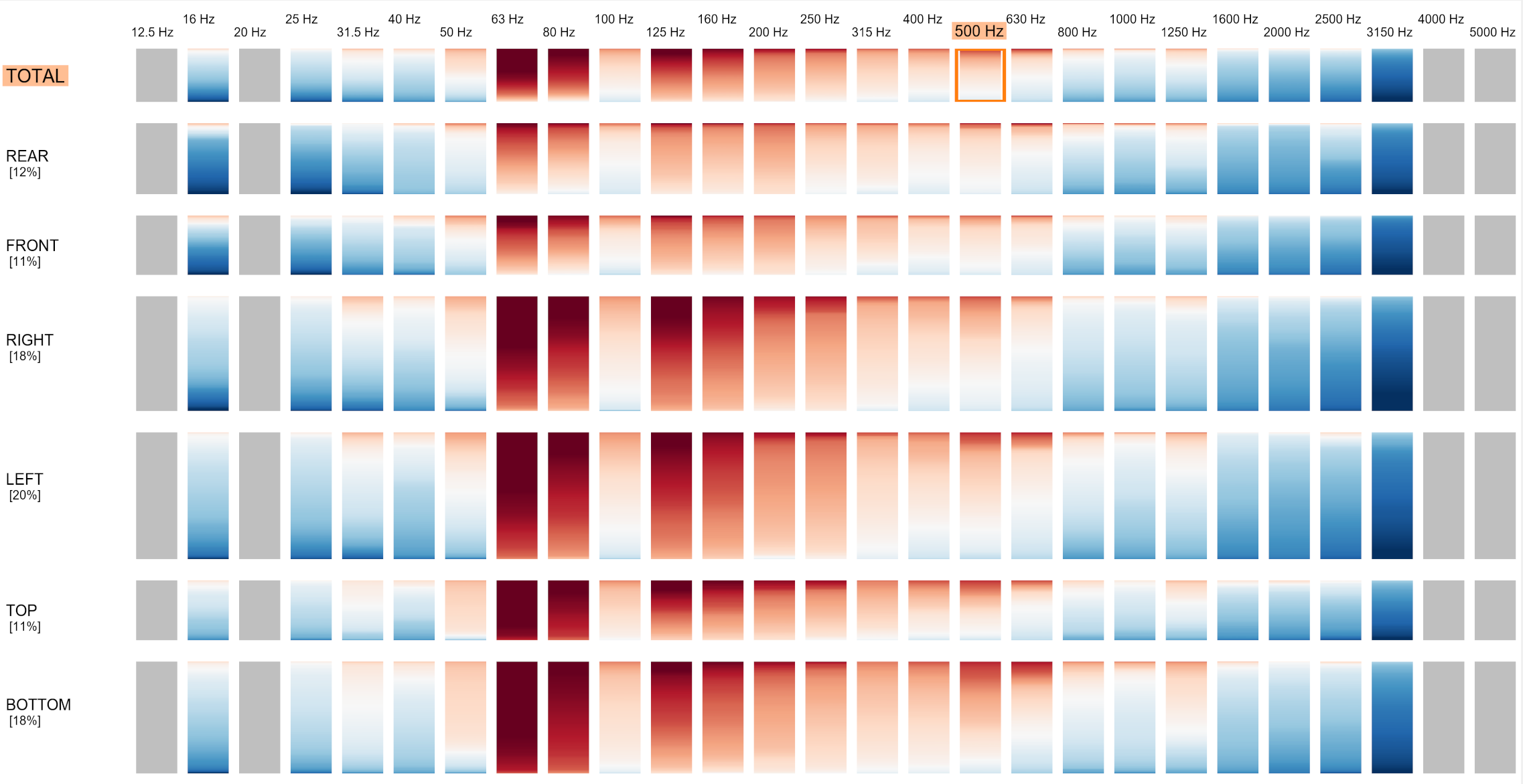}
    \caption{Medium-level analysis using area weighted and ranked discrete velocity levels using a diverging color scale~-- blue\,$\leftrightarrow$\,60\,dB, white\,$\leftrightarrow$\,90\,dB, red\,$\leftrightarrow$\,120\,dB.}
    \label{fig:cell_velocity_levels_wrt_colored}
\end{subfigure} \hfill

\caption{The overview matrix can be shown in different modes to support overview and medium level tasks.}
\label{fig:cell_velocity_levels:all}    
\end{figure} 
\hfill

%
%
%
%
\subsection{Providing the necessary overview}
\label{sec:design_overview}
As described in Section~\ref{sec:tasks}, the goal of task T1 is to gain a high-level overview across frequency bands and engine parts with a special focus on whether the design fulfills the externally prescribed thresholds.  
To quickly gain such an overview at three levels for all frequency bands and engine parts (requirement R1), we found a 2D matrix layout appropriate with engine parts as rows and frequency bands as columns.
The first row represents the same data as shown in one row of a Campbell diagram, i.e., a single value per frequency (band) for the whole engine for a specific operation point. The following rows represent partitions of the engine and add spatial domain information which can not be found in a Campbell diagram.
The row height corresponds to the size of the corresponding partition.
Since noise regulations specify the range which is considered borderline and \replaced{since}{thus} a value \added{thus} falls into one of three acceptance categories (achieved, borderline, not achieved), we decided to use the traffic light metaphor \added{for visualization}. Engineers are familiar with this concept and immediately get an effective overview. 
For users with a different social and/or cultural background\added{,} this color scheme can be \replaced{adapted}{easily exchanged}.  
In our case, we used a colorblind-friendly color scheme that maintains the connotation that the equivalent of the red hue is the `bad' side and a green hue for the `good' side. Simulated frequency bands, for which no thresholds are prescribed, are shown in gray ({Fig.~\ref{fig:overview_limits}}).  
The top row shows the overall noise levels per frequency band, integrated over all engine parts~-- should be ``all good''.  
%
\subsection{Prioritizing parts for in-detail examination}
Given that a design is not fulfilling all prescribed thresholds, 
 engineers need to start a physics-oriented analysis in more detail.  
To do so, they need to first prioritize\deleted{,} which parts of the design they must check in more detail (R2).  
A more detailed depiction of noise levels, and their distribution across harmonics, is needed as well as an indication of the engine parts' sizes. 
As it is still necessary to show this information across all frequency bands and engine parts, we agreed to keep the same matrix layout and size\deleted{ consistent}. Hence\added{,} the part sizes are already encoded in the overview level \replaced{and}{but} do not impair its efficacy.
\par 
In order to prioritize critical parts based on the externally prescribed levels, a refinement of the overview mode\added{, as} introduced in Section~\ref{sec:design_overview}\added{,} is used. In this refinement, we show  the single scalar acceptance values by using a coloring mode that is inspired by two-tone pseudo coloring~\cite{TwoTone}, helping engineers to quickly estimate the distance to the neighboring acceptance categories  (Fig.~\ref{fig:overview_limits_horizon}).
The actual integral velocity level acceptance value is encoded in such a way that ratio of the two shown tones indicates the closeness of the value to the respective acceptance category which is defined per frequency band. The height of the grid cell maps to the size of the borderline range of 6\,dB and the middle of the grid cell can be interpreted as the position of the actual value within the range\replaced{ (}{. }Fig.~\ref{fig:limits_encoding}\added{)}.
 It is also possible to subdivide the acceptance categories in the vicinity of the borderline range into a configurable number of corresponding color shades by changing the lightness and keeping the tone/hue constant. The additional shades ease the quantitative assessment of the limits violation severity, e.g., when using three shades one shade represents a range of 2\,dB. Fig.~\ref{fig:overview_limits_horizon} illustrates how the encoding works without and with additional shades.

For a more detailed overview the computed velocity levels themselves can be shown in the matrix view.
In this mode the velocity values are shown as ranked values of all surface elements that belong to a partition (Fig.~\ref{fig:cell_velocity_levels_wrt_colored}).
Another mode shows these ranked discrete levels using the traffic light metaphor {Fig.~\ref{fig:overview_discrete_limits}} described in section {Section~\ref{sec:design_overview}}.
Since all these modes are displayed alternatively in the matrix layout an additional mode shows integral and discrete velocity level acceptance concurrently {Fig.~\ref{fig:overview_combined}}.
Hence the user can choose to switch between the modes for integral and discrete velocity levels or show them concurrently\deleted{ depending on her preferences}.
The different modes help  engineers in prioritizing problematic elements. 
The top-level mode remains important for quick checks at the overview level, while the second level modes require more attention, but provide relevant additional details. 
%
\begin{figure}[t!]
\centering
\begin{subfigure} {0.49\columnwidth}
    \centering
\includegraphics[width=\columnwidth]{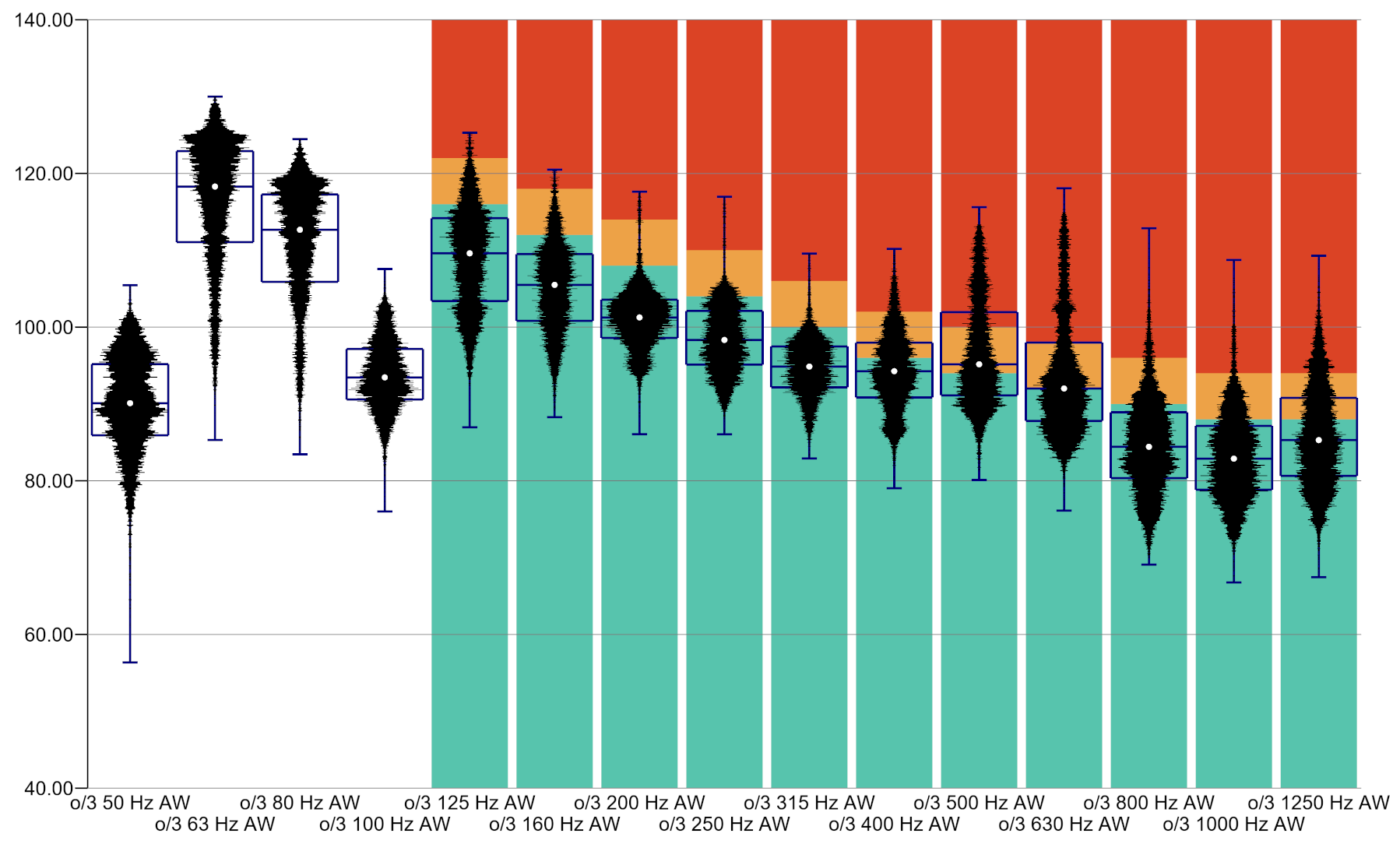}
    \caption{}
    \label{fig:boxplot}
\end{subfigure}
\begin{subfigure} {0.49\columnwidth}
    \centering
\includegraphics[width=\columnwidth]{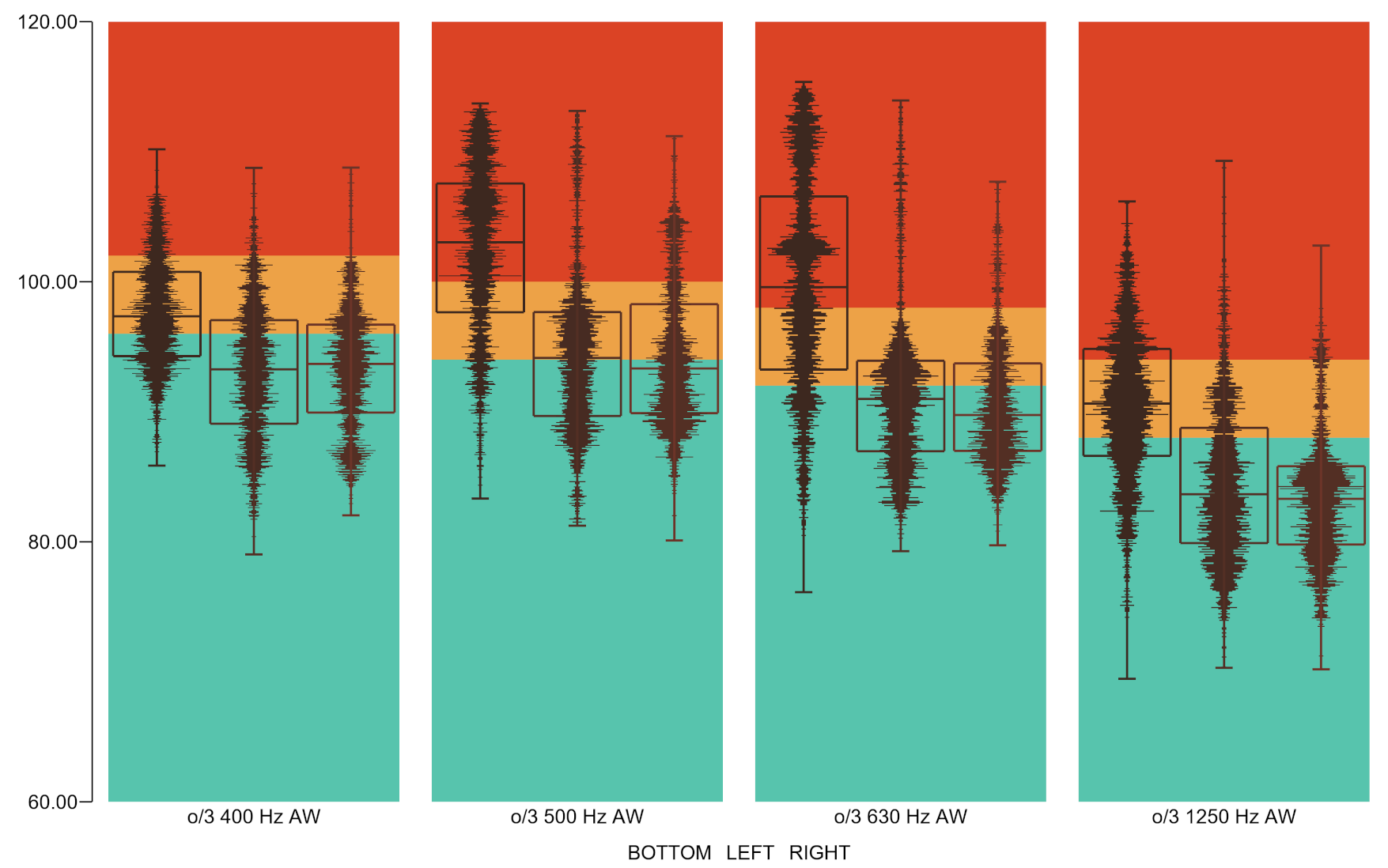}
    \caption{}
    \label{fig:boxplot_split}
\end{subfigure}
\caption{Using boxplots to show discrete velocity level distributions against discrete velocity level acceptance categories without (a) and with split (b) per specified engine parts (BOTTOM, LEFT, RIGHT). When split\added{,} an individual distribution is shown for each specified part inside each of the bars representing the specified \nicefrac{1}{3} octaves.} 
\label{fig:boxplots}
\end{figure}
\begin{figure}[t!]
\centering
\begin{subfigure} {.49\columnwidth}
    \centering
    \includegraphics[width=\columnwidth]{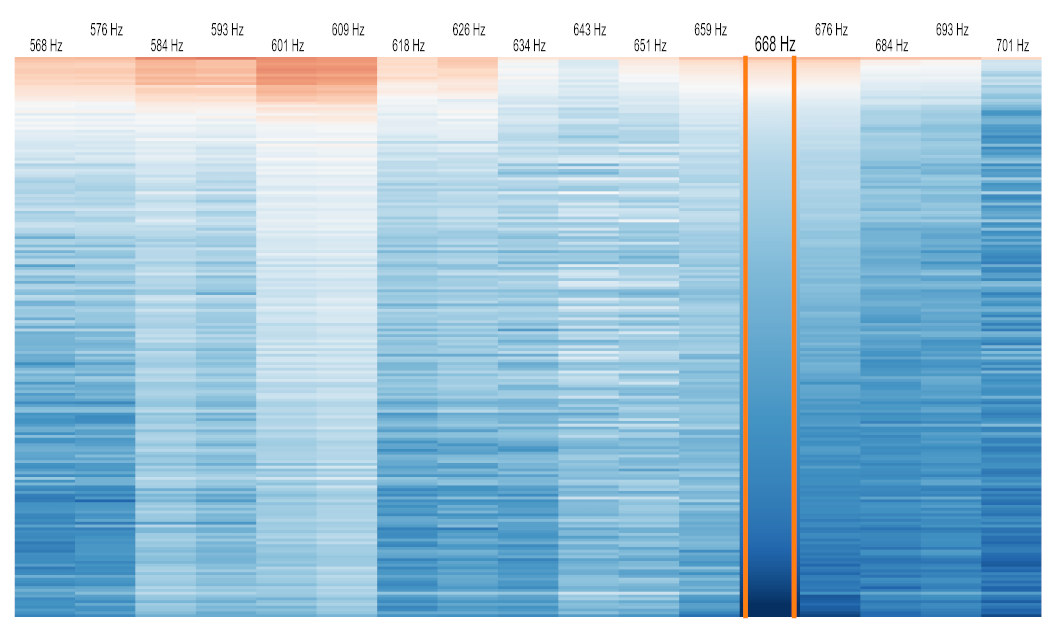}
    \caption{sort by selection}
    \label{fig:total_630Hz_harmonics_single}
\end{subfigure} \hfill
\begin{subfigure} {.49\columnwidth}
    \centering
    \includegraphics[width=\columnwidth]{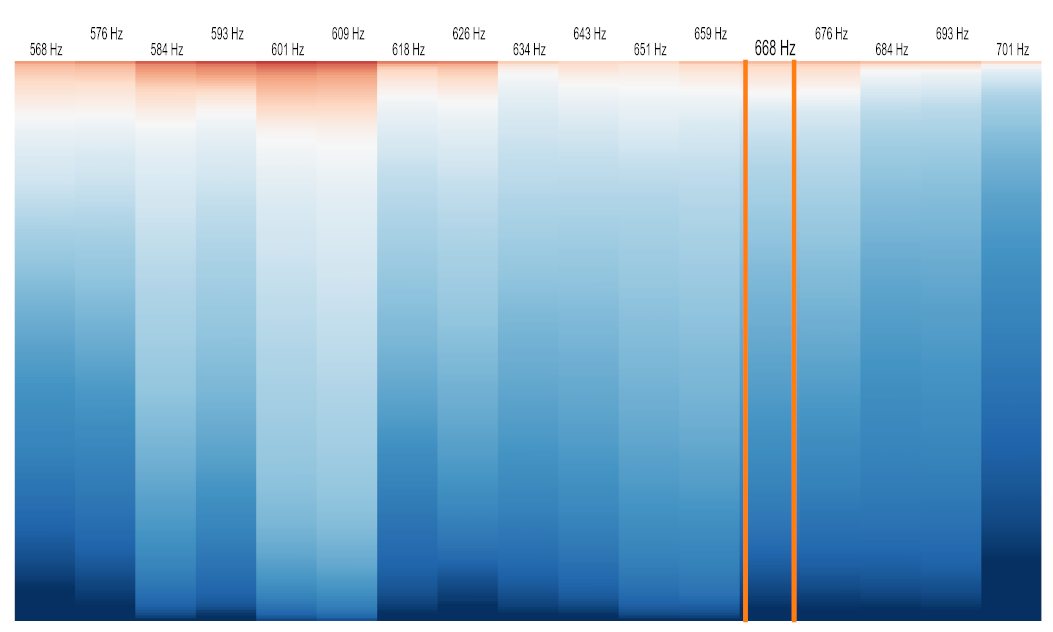}
    \caption{sort all individually}
    \label{fig:total_630Hz_harmonics_all}
\end{subfigure} \hfill
\caption{Discrete velocity levels of harmonics at the 630\,Hz \nicefrac{1}{3} octave of region TOTAL, shown according to different sorting approaches. The column height represents the area of the selected region and is subdivided into a specified number of equally sized rows. Hence multiple cell values can contribute to the value of one row. In this case the maximum value of all contributing cells is used.
Color scale: blue\,$\leftrightarrow$\,60\,dB, white\,$\leftrightarrow$\,90\,dB, red\,$\leftrightarrow$\,120\,dB.}
\label{fig:total_630Hz_harmonics}
\end{figure}
\begin{figure*}[t!]
\centering
\includegraphics[width=\textwidth]{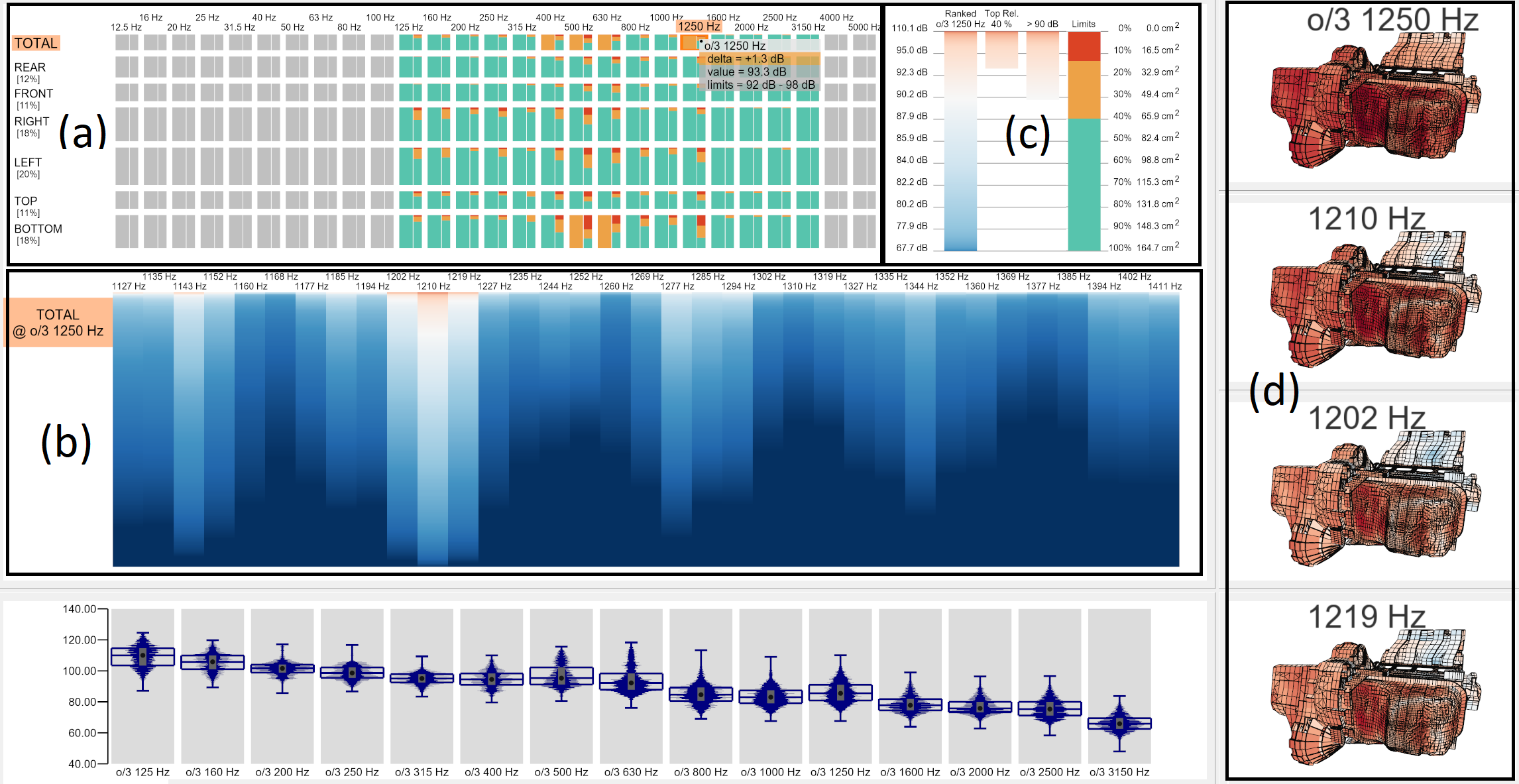}
\caption{The drill-down view consists of three panes: overview matrix pane (a), harmonics pane (b), and frequency band details pane (c). It is the central hub for the exploration process. The three most critical harmonics of \nicefrac{1}{3} octave 1250\,Hz are automatically shown in 3D views on the right (d) as \nicefrac{1}{3} octave 1250\,Hz is selected in the drill-down view.} 
\label{fig:geo3d_multi}
\end{figure*}
%
%
%
\par 
A series of \replaced{boxplots}{box-plots} with integrated histograms provide a better overview of the frequencies distributions across the cells, and still scales well with the number of cells. In contrast to the drill-down-view, which uses color encoding of the velocity level values, the \replaced{boxplots}{box-plots} uses positional encoding.
Fig.~\ref{fig:boxplot} shows \replaced{boxplots}{box-plots} with a mirrored histogram for each of the frequency bands of interest. The background stripes show the limits, so that engineers can see how many cells belong to each level. In order to not loose the spatial context we suggest to split the plots according to the engine parts (Fig.~\ref{fig:boxplot_split}). 
\subsection{Detailed analysis across both domains}
%
%
%
As soon as potentially problematic design parts are prioritized for a detailed examination, additional perspectives and interactions become necessary (requirements R3, R3a, and R3b).  
We need to show the velocity levels for all selected cells and for all harmonics that belong to the selected frequency band (R3a).  
To maintain a consistent coding, we apply the same color scheme as for the intermediate level, but additional visualization space is needed to account for the usually large number of cells.  
Since showing all harmonics is not an option, we decided to show all  harmonics that belong to a selected frequency band (1 to 69 harmonics per frequency band).  
We dedicate one vertical bar to each harmonic and color it according to the area-normalized values of the discrete velocity levels.  
We place these bars right next to each other to optimize the use of the available visualization space.  This provides a good way to compare multiple harmonics to each other.  
%
\par 
To support all relevant analysis tasks, the view provides two sorting strategies:~ 
If each column is sorted independently, a quick overview of the area size and the magnitude of the velocity levels is provided. 
In this case, the lines in one row can belong to different cells (Fig.~\ref{fig:total_630Hz_harmonics_all}).~ 
If spatial coherence is required, the user can sort the entire view by one selected column.  
This way, the engineer sees which harmonics share similar characteristics (Fig.~\ref{fig:total_630Hz_harmonics_single}). 
%
%
%
%
\par
\begin{figure}[t!]
\centering
\includegraphics[width=.8\columnwidth]{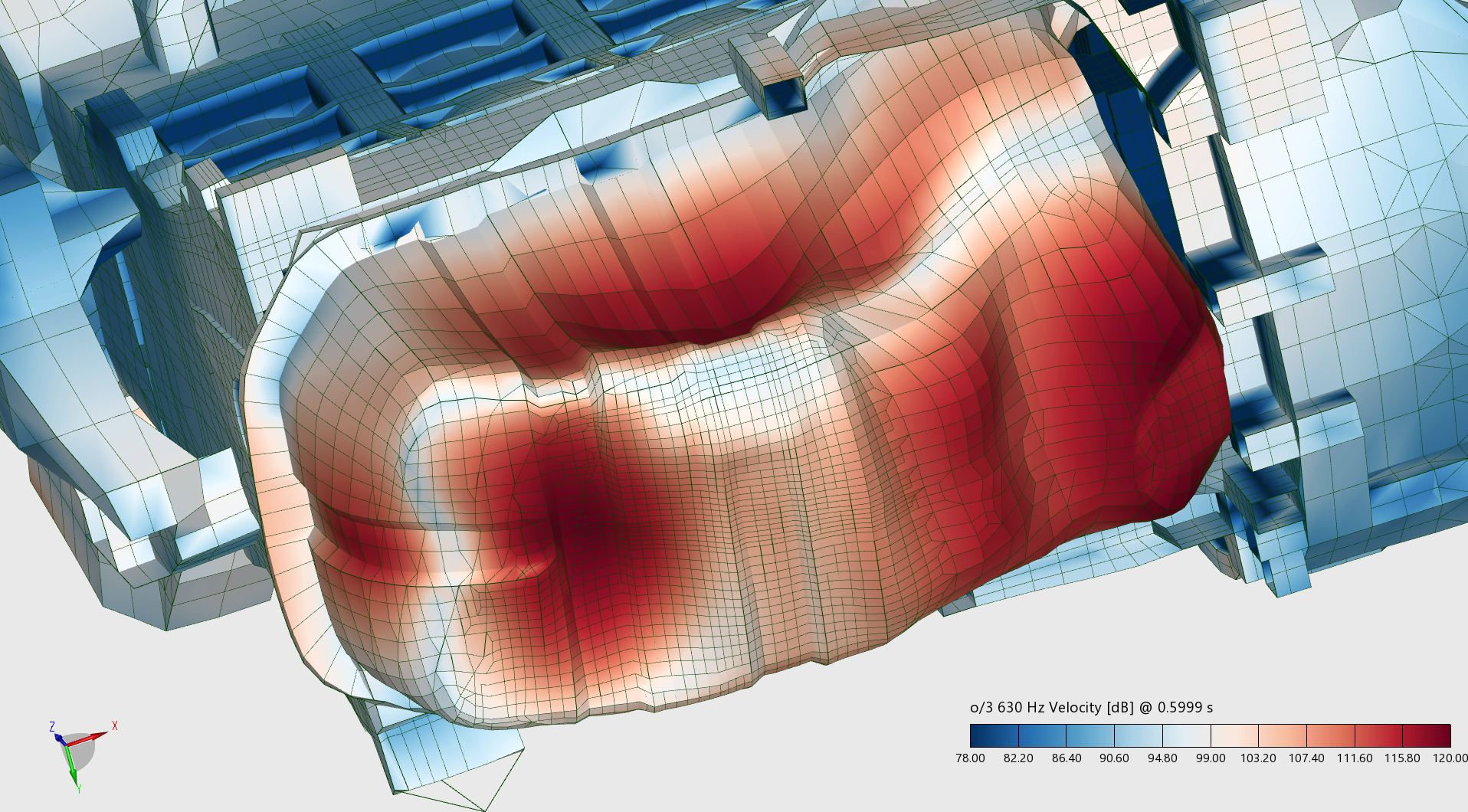}
\includegraphics[width=.8\columnwidth]{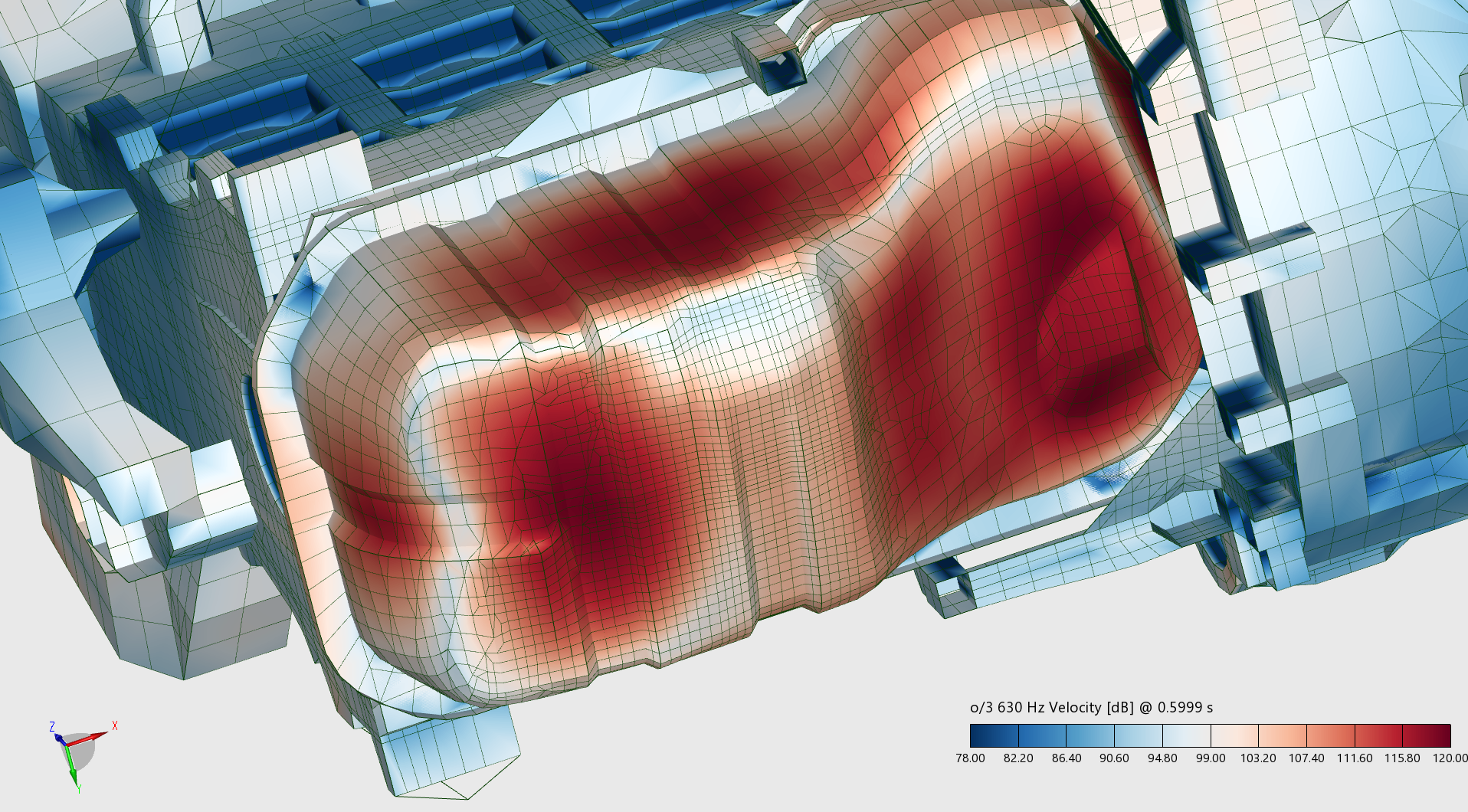}
\caption{Oil pan with an\added{d} without exaggerated deformation at the 630\,Hz \nicefrac{1}{3} octave.} 
\label{fig:geo3d_deformation}
\end{figure}

 
To achieve a more detailed analysis in the spatial domain (R3b), we integrate a suitably adapted 3D geometry view.
Consistently, we also use the same color coding in this view to show velocity levels.  
To compare velocity levels across several harmonics that belong to the same frequency band, a series of linked 3D views is used~-- changing the  perspective in one view leads to an according change in all other views (Fig.~\ref{fig:geo3d_multi}).  
The 3D view is also important to judge radiation efficiency (T3c). 
Velocity levels are indicators of noise, but the same velocity level can cause different levels of noise.  
If an engine part swings without changing its volume at a certain velocity, the resulting noise will be low. 
If, however, a part changes its volume, the noise will be high. 
Engineers need to inspect these parts and, based on their knowledge and experience, they can judge which parts with high velocity levels are indeed critical.  
An animated 3D view showing exaggerated deformations helps to identify spots with high radiation efficiency (Fig.~\ref{fig:geo3d_deformation}). 
\par 
Furthermore, engineers need to detect hot-spots at the surface (T3d). 
For the color mapping we opted for diverging shades instead of sequential scales to let the user see more differences in the data and to emphasize the extremes.
Engineers are also familiar with this kind of color mapping.
Custom color scales proved to be very efficient to achieve this. 
We decided to use a non-linear color scale which compensates for the  logarithmic effects in the data given in decibels.  Fig.~\ref{fig:400_non_linear} shows this color mapping and how it supports the detection of hot-spots at 400\,Hz.  
%
\par 
As we need to provide a joint analysis of both frequencies and spatial locations (R3), we found linked views a suitable solution.  
As soon as the user selects a harmonic in the harmonics pane, the selection is also shown in the linked 3D view, and vice versa. 
\subsection{Interaction design}

\begin{figure*}[t]
\centering
\includegraphics[width=\textwidth]{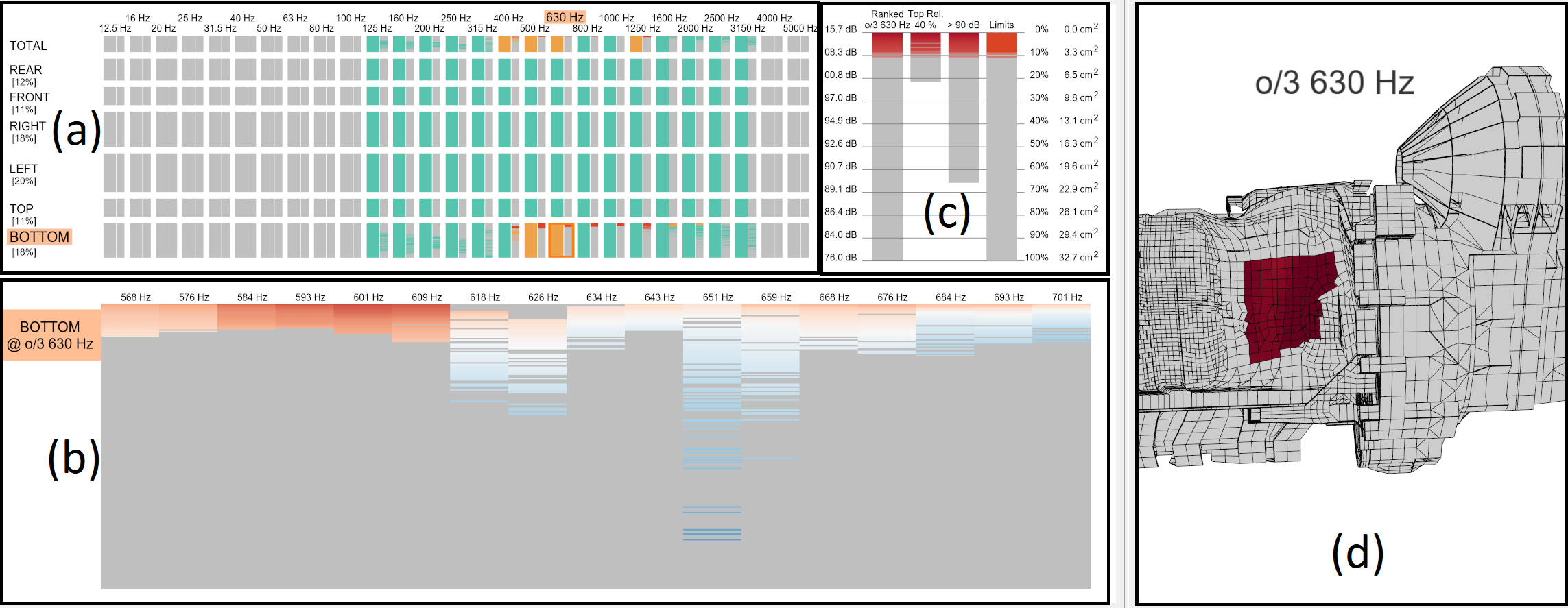}
\caption{Cells selected in \added{the} spatial domain in the 3D geometry view (d) are highlighted in \added{the} frequency domain in each of the panes (a), (b), and (c) of the drill-down view.}
\label{fig:geo3d_selection}
\end{figure*}

NVH data is multi-faceted and it is not feasible to show all data in one integrated visualization.  Linked views with appropriate interaction mechanisms allow to integrate several perspectives in the  analysis~\cite{Keim2008}.  
We therefore link all views so that all selections and changes in any one view immediately get reflected consistently in all other views.  In the following, we discuss our interaction design, related this setup of linked views.  
%
\par 
Starting from the overview, engineers need to drill down into prioritized frequency bands and engine parts.  
To aid the detailed data assessment, we automatically show the corresponding harmonic pane ({Fig.~\ref{fig:total_630Hz_harmonics}}) when the user hoovers the pointer over a rectangle in the overview.  The user can also click on any field in the overview to ``freeze'' the selection.  
\par 
To refine the selection, the user can interact with the frequency band details pane (Fig.~\ref{fig:geo3d_multi}(c)), showing the integrated velocity values for the selected frequency band in four bars.  
The scales are linear in area size, as surface areas are key to  identifying critical places.  
The right-most bar shows the proportion of threshold areas, i.e., the discrete velocity level acceptance. It is a magnified version of a particular cell shown in Fig.~\ref{fig:overview_limits_horizon}. Clicking on any of the three parts refines the selection accordingly.  
The third bar provides access to a subset of cells with velocity levels that exceed a certain, user-defined threshold.  
The second bar shows the most critical cells using a relative, percentage-based threshold, while the left-most bar is showing all values, ranked by the velocities.  
As the layout is linear with respect to the cell areas, the corresponding decibel values spread out according to their distribution.  
Grid lines help with related the bars to their respective scales.  
\par 
It is also important to let users select cells in the 3D views, for example, by clicking onto one or several of them.  
A particularly useful interaction is to enlarge the set of selected cells, including additional neighboring cells.  
We found it helpful to also provide a growing mode that only includes additional cells if they exceed a certain velocity level.  
This way, even larger areas of relevant values can be easily selected in the 3D view (Fig.~\ref{fig:geo3d_selection}).  
Of course, standard interactions like zoom and rotation are also supported.  
Importantly, the 3D views can also be linked to always show the same perspective, regardless which of the views the user is interacting with.
%
\par 
We integrate the overview matrix pane, the harmonics pane, and the frequency band details pane into a single drill-down view with three panes.  
These visualizations are always used and represent the central point in the exploration process.
After several iterations in our collaborative design process, we arrived at the decision to keep this view~-- the drill-down view~-- as a stable component in the top-left of the visualization at all time points.  
The drill-down view is integrated with additional coordinated views, including and adapted \replaced{boxplots}{box-plots} and geometry views.  
Fig.~\ref{fig:geo3d_multi} shows a typical configuration from one of our evaluation sessions with 3D views showing the selected frequency band and three most critical harmonics right next to the drill-down view, as well as the adapted \replaced{boxplot}{box-plot} view underneath.
\section{Use Case}
\label{sec:evaluation}
%
%
To evaluate our approach, we engaged with a case study of data from a  multi-body dynamics (MBD) simulation of a front-wheel drive (FWD) demo model at an operation point of 2000\,RPM using AVL EXCITE\texttrademark{}, a tool for advanced durability and NVH analysis.  
The model covers an inline four-cylinder (I4) internal combustion engine, a manual-transmission gearbox, the final drive/differential, and front axle shafts with brake flanges. The model is set to the 3\textsuperscript{rd} gear. 
Three mechanical engineers who also coauthor the paper conducted the evaluation. \replaced{Below, w}{W}e describe the data first, \replaced{followed}{following} by the evaluation.

\subsection{Simulation data}
\label{sec:sim_data}
\begin{figure}[t!]
\centering
\includegraphics[width=.4\columnwidth]{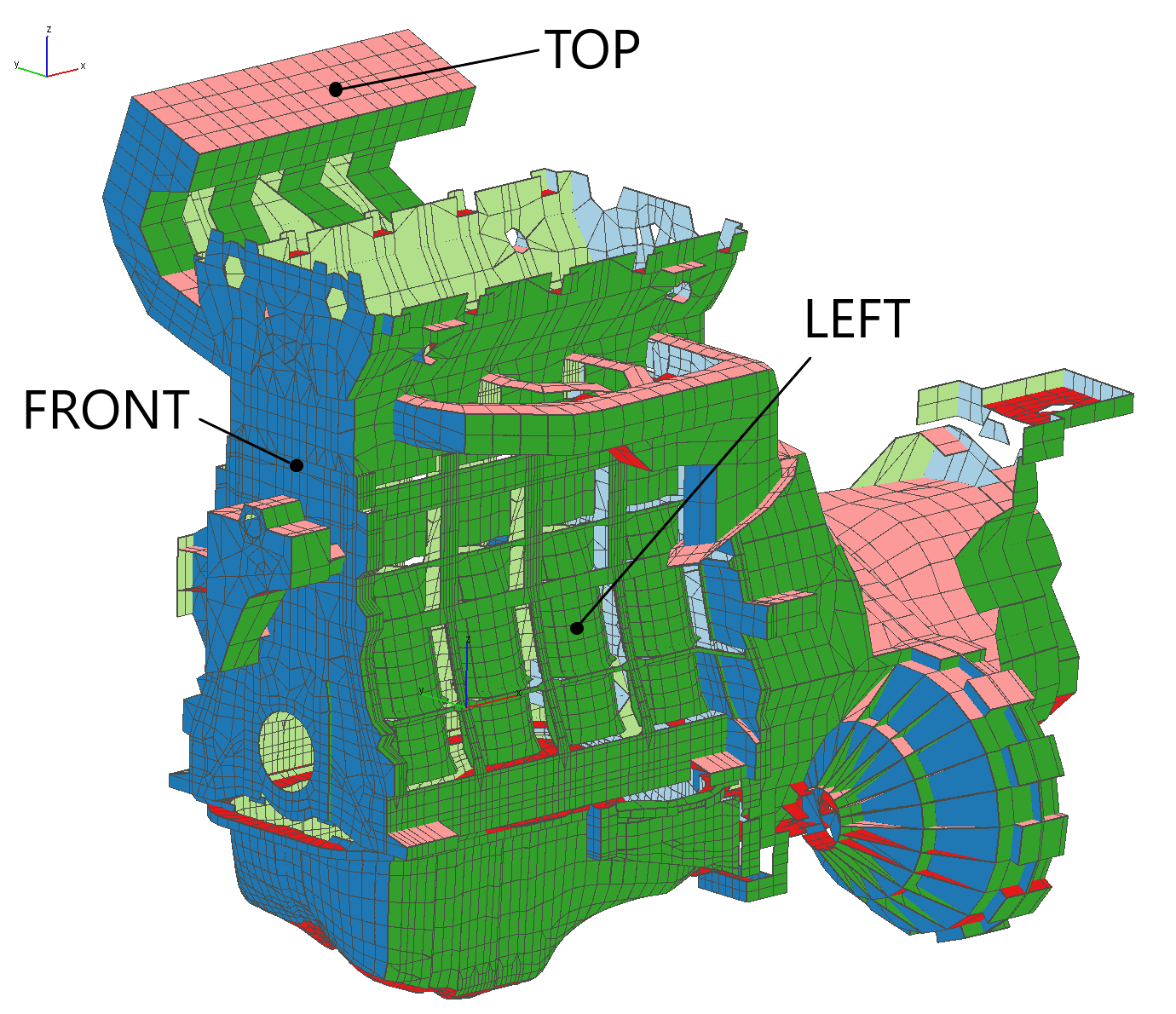}
\includegraphics[width=.4\columnwidth]{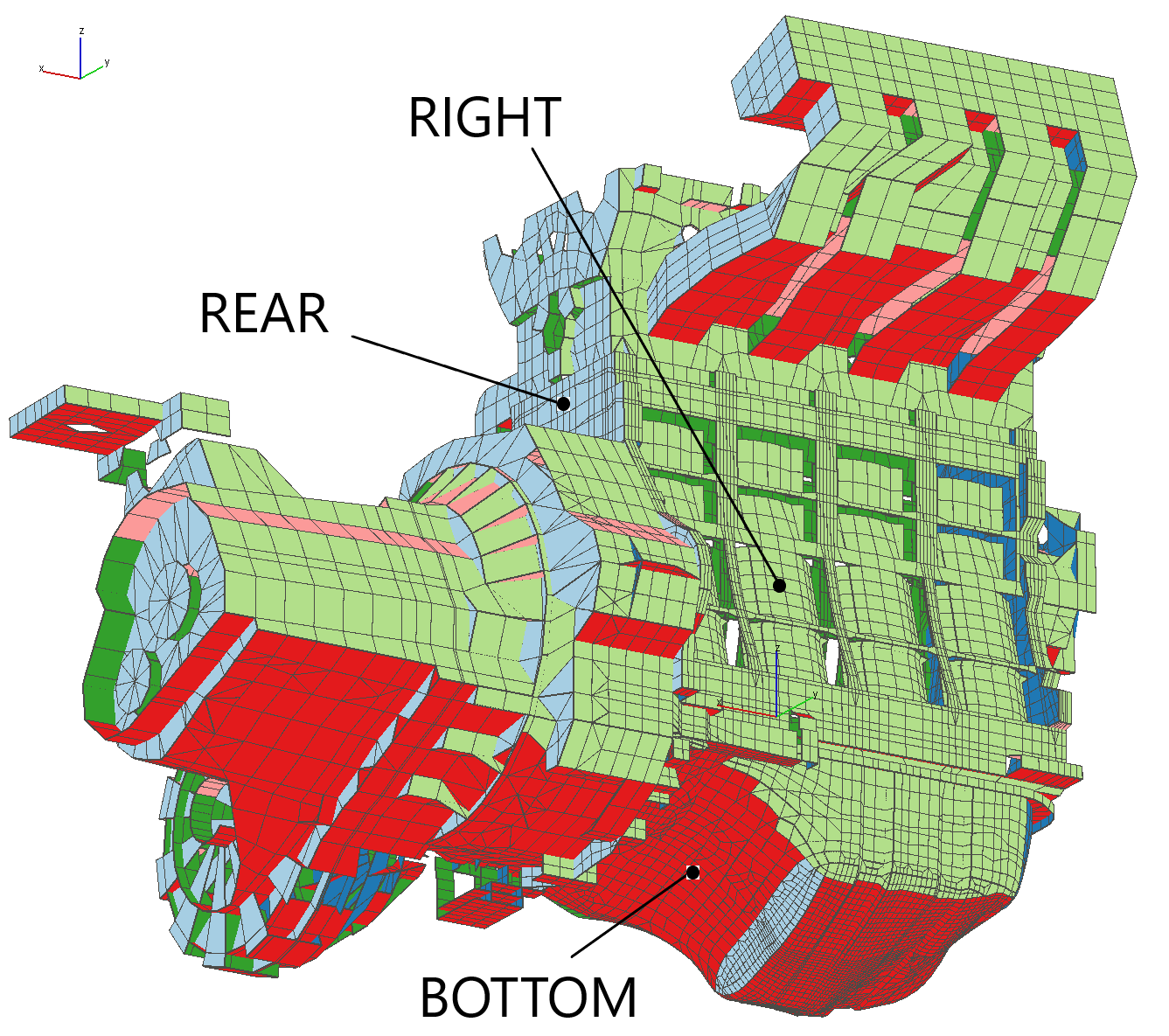}
\caption{Outer surface partition of the power unit.}
\label{fig:selections_front_left_top}
\end{figure}

Simulation and data recovery result in kinematic quantities (for displacement, velocity, and acceleration) in \added{the} time domain for the surface of the model.
Nodal kinematic data is calculated in \added{the} frequency domain from which the average normal-to-surface velocity of each outer cell is obtained.  The velocities are normalized to a reference area of 1\,m\textsuperscript{2} and expressed in dB using AVL's standard reference \deleted{velocity }value.  
\par 
In this case study, the outer surface has 13\,047 vertices and 12\,173 cell faces and it is partitioned into engine sides FRONT, REAR, TOP, BOTTOM, LEFT, and RIGHT (Fig.~\ref{fig:selections_front_left_top}).  
Integrating the cell faces' normal velocities across each surface partition area and/or each standard frequency band (octave, \nicefrac{1}{3} octave)\added{,}  we get the data as analyzed in this use case: 
\begin{itemize}
\itemsep=0pt
\item
For each cell face at the outer surface we get the sub-region it belongs to, its area, and 
the discrete velocity level in dB for each of 359 harmonics and 26 \nicefrac{1}{3} octave\deleted{  frequency band}s.
\item
For each surface region we get its area and the integral velocity level for each harmonic and \nicefrac{1}{3} octave\deleted{ band}.
\end{itemize}
%
%
%
%
The simulation results are checked against two given criteria for velocity levels (both prescribed for \nicefrac{1}{3} octave\deleted{ band}s in the 215\,Hz\,\,--\,\,3150\,Hz range at 2000\,RPM): 
\begin{itemize}
\itemsep=0pt
\item
Integral velocity level limits in dB, applicable to levels obtained by integration of discrete levels across the whole outer surface of an engine.
\item
Discrete velocity level limits in dB, applicable to velocity levels calculated for the cells.
\end{itemize}

\begin{figure*}[t!]
\centering
\begin{subfigure} {0.3\textwidth}
    \centering
    \includegraphics[width=\columnwidth]{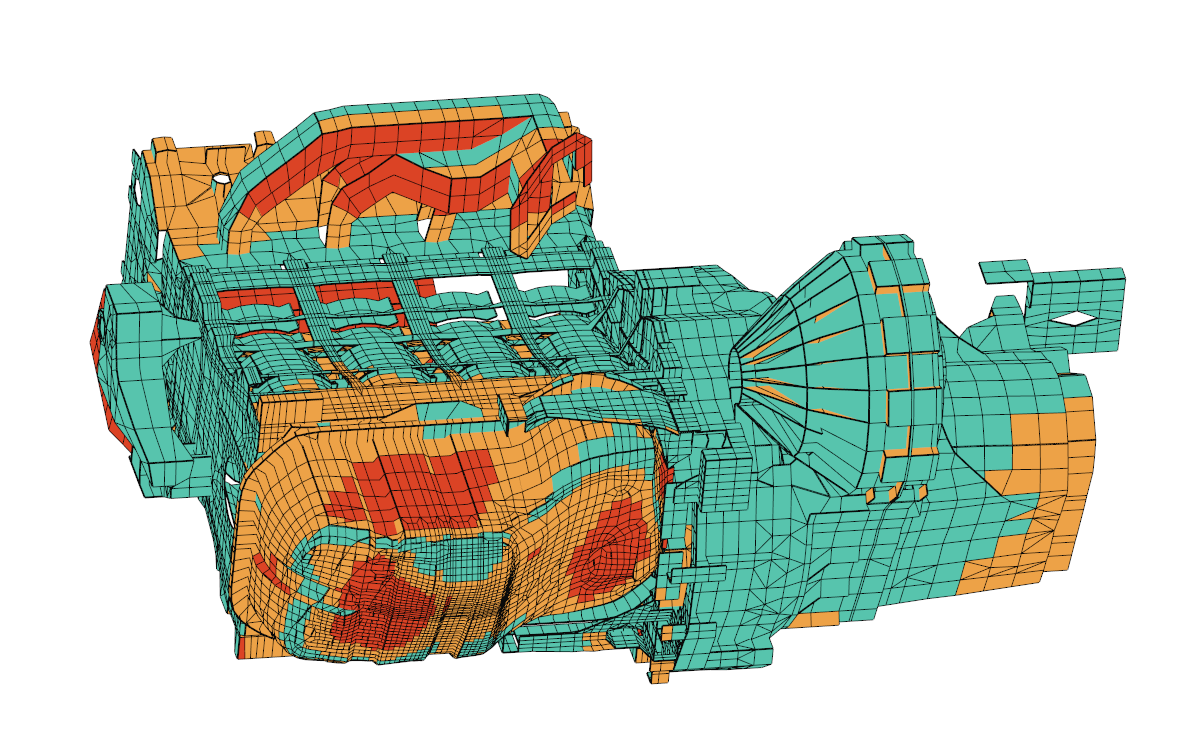}
    \caption{Limits}
    \label{fig:400_limits}
\end{subfigure} \hfill
\begin{subfigure} {0.33\textwidth}
    \centering
    \includegraphics[width=\columnwidth]{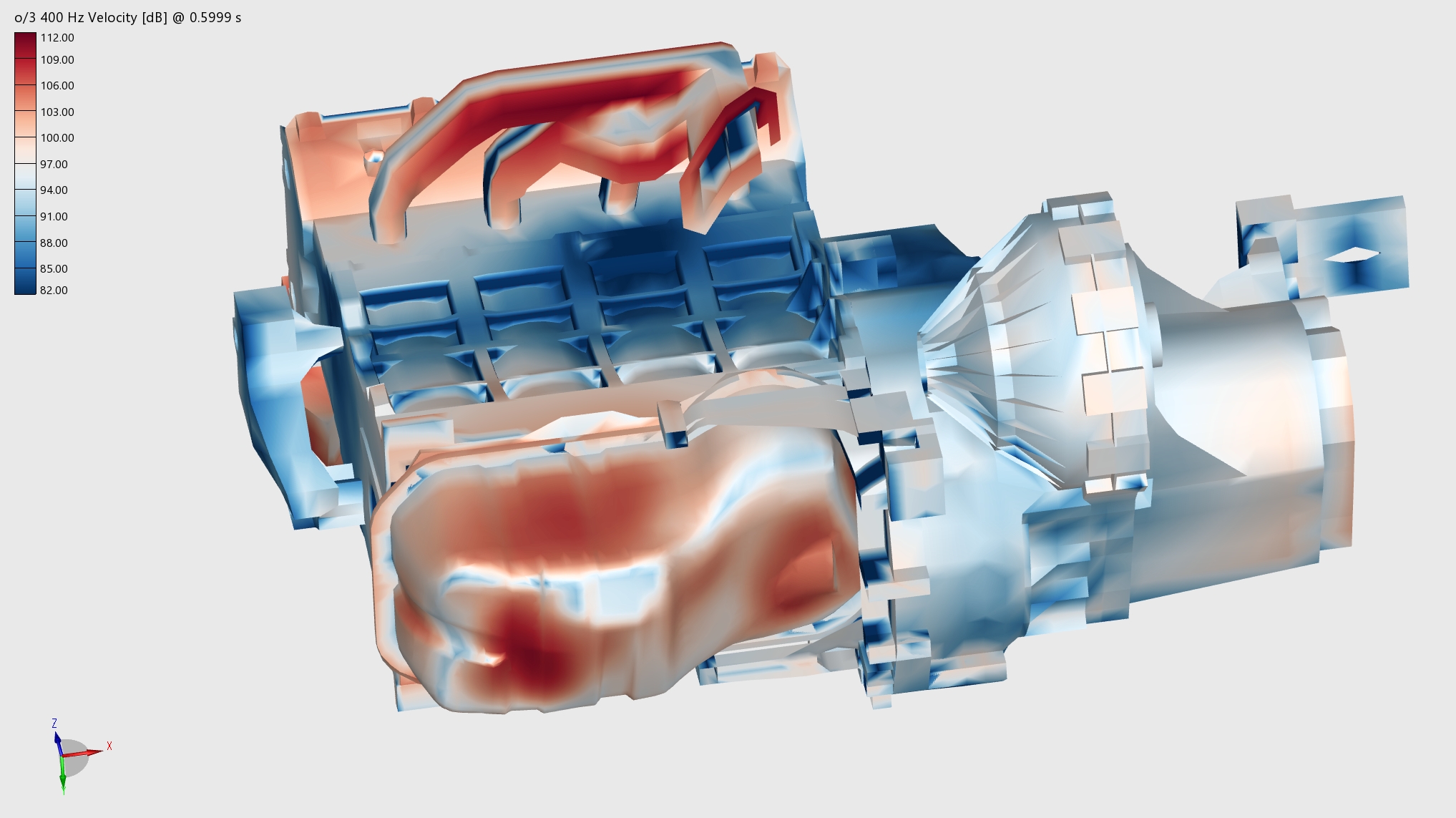}
    \caption{Linear}
    \label{fig:400_linear}
\end{subfigure} \hfill
\begin{subfigure} {0.33\textwidth}
    \centering
    \includegraphics[width=\columnwidth]{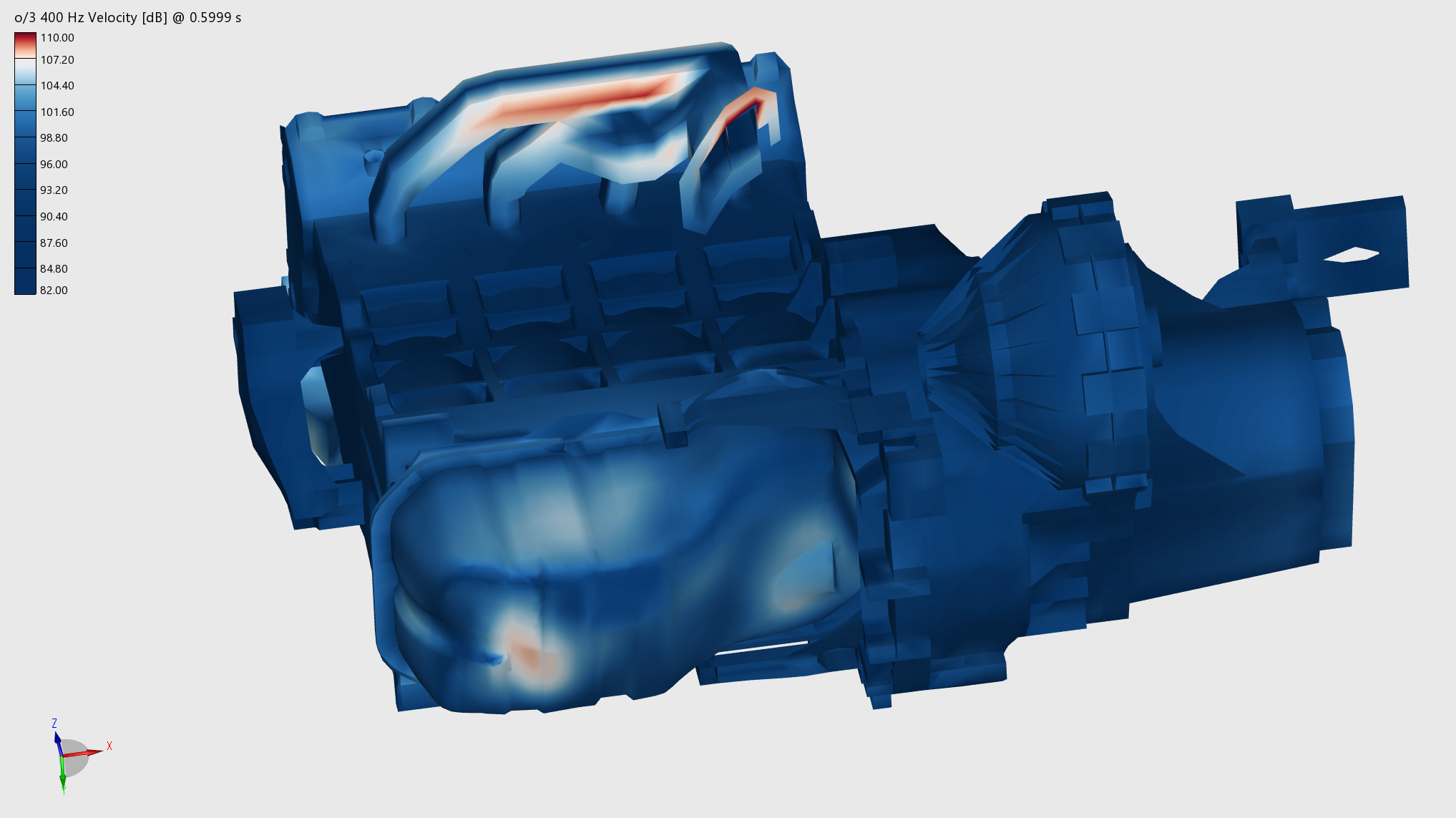}
    \caption{Non-linear}
    \label{fig:400_non_linear}
\end{subfigure}
\caption{Visualization of velocity levels \replaced{at the}{@} 400Hz \nicefrac{1}{3} octave with three different color mappings.}
\label{fig:limits_linear_non_linear}
\end{figure*}

\subsection{Overview first}

The first task in the exploration process is gaining overview on parts and frequencies that are not conform to the prescribed limits (Fig.~\ref{fig:overview_limits}). 
Using the overview matrix pane of the drill-down view, the overall engine ``pass criteria'' is checked.  
The yellowish rectangles in the TOTAL row for \added{the} 400, 500, 630 and 1250\,Hz \nicefrac{1}{3} octaves indicate frequency bands where values are above \added{a} threshold\deleted{s}.

%
\par 
A non-greenish color (yellowish or reddish) of rectangles related to  sub-regions warns the analyst of high velocity levels in these sub-regions and within the corresponding frequency bands.  
It shows that these sub-regions~-- at these frequency bands~-- contribute the most to the engine's overly high integral velocity levels.  
In the shown example, it is the BOTTOM region at the 500 and 630\,Hz \nicefrac{1}{3} octave.
The BOTTOM region contains the oil pan, a part of the engine which often requires a special attention \replaced{during}{when} noise analysis\deleted{ is in focus}.  
\par  
Subsequently, the severity of the acceptance limits violation for the identified critical frequency bands at the engine level (represented by TOTAL region/row) is checked 
{qualitatively} by switching to \replaced{our}{the} two-tone pseudo coloring mode
(Fig.~\ref{fig:integral_velocity_level_limits_horizon_shadows}).  
This visualization mode shows how ``deep'' we are within the corresponding acceptance category (in our case: the yellowish ones).

\subsection{Prioritize}

\begin{figure}[t!]
\centering
\includegraphics[width=\columnwidth]{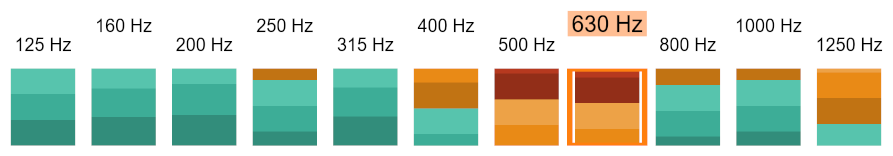}
\caption{Engine acceptance shown with \added{our} two-tone pseudo coloring using three shades for \deleted{each of} the acceptance category tones (TOTAL region).}
\label{fig:integral_velocity_level_limits_horizon_shadows}
\end{figure}

Once the overview is gained, the experts prioritize the problematic parts in order to solve the biggest problems first.
Comparing frequency bands with each other in Fig.~\ref{fig:integral_velocity_level_limits_horizon_shadows}, we also see relative violation severity, providing an importance order, in which the critical regions\,/\,frequency bands should be examined: in the 400\,Hz \nicefrac{1}{3} octave we exceeded the acceptance limit just slightly (shades of green), in the 1250\,Hz \nicefrac{1}{3} octave we entered more into the borderline area (shades of yellow), while in the 500 and 630\,Hz \nicefrac{1}{3} octaves we are \deleted{quite }close to the unacceptable region (shades of red). 
So, the latter two frequency bands are examined first.  

\begin{figure}[t!]
\centering
\includegraphics[width=.55\columnwidth]{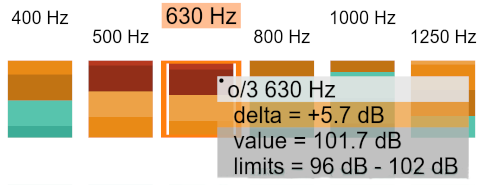}
\caption{Quantified severity of the acceptance limits violation.}
\label{fig:velocity_levels_distributions}
\end{figure}

%

%
%
%
\begin{figure*}[t!]
\centering
\begin{subfigure} {0.33\textwidth}
    \centering
    \includegraphics[width=\columnwidth]{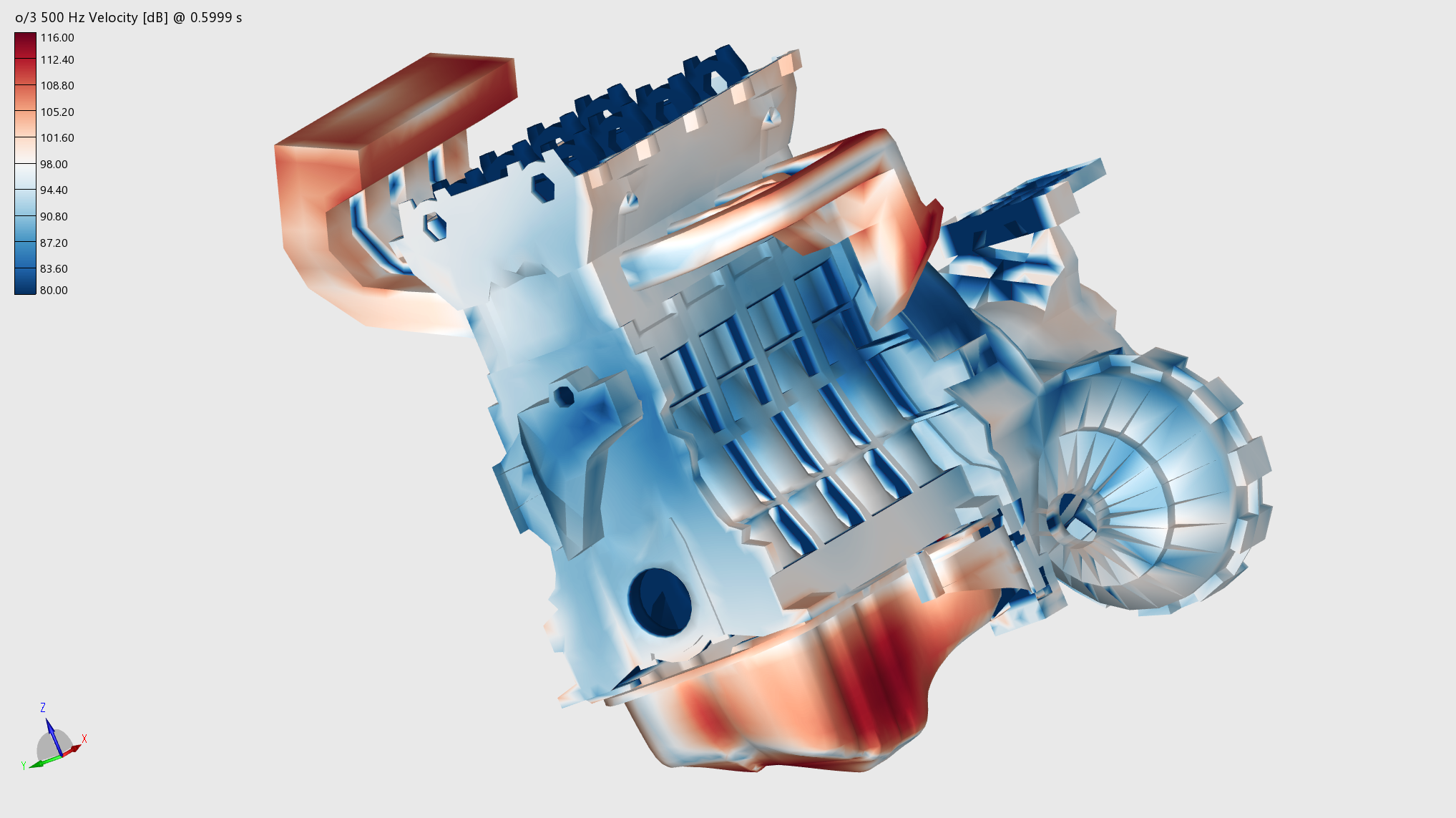}
    \caption{500Hz \nicefrac{1}{3} octave}
    \label{fig:500_linear}
\end{subfigure} \hfill
\begin{subfigure} {0.33\textwidth}
    \centering
\includegraphics[width=\columnwidth]{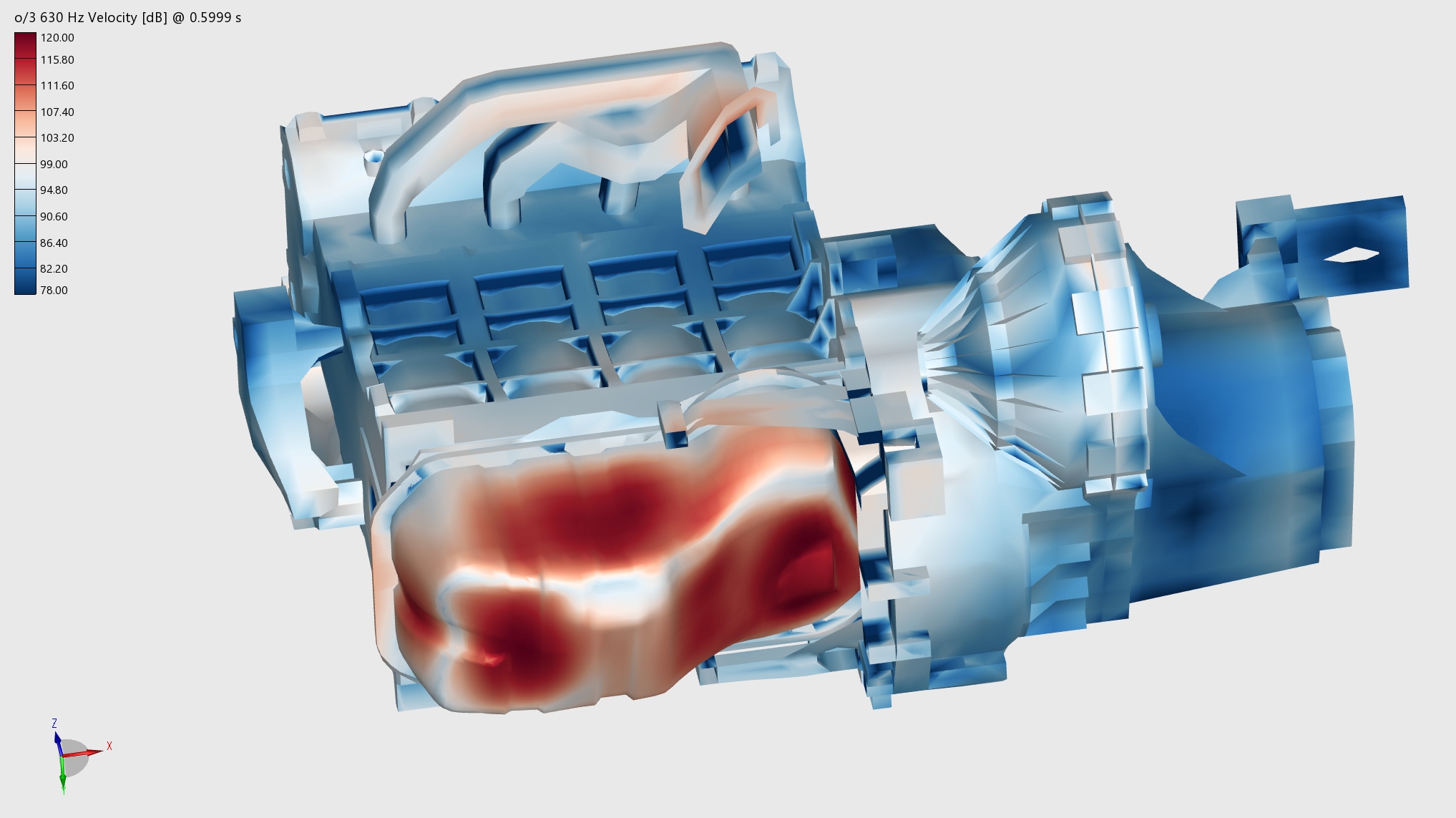}
    \caption{630Hz \nicefrac{1}{3} octave}
    \label{fig:630_linear}
\end{subfigure} \hfill
\begin{subfigure} {0.33\textwidth}
    \centering
    \includegraphics[width=\columnwidth]{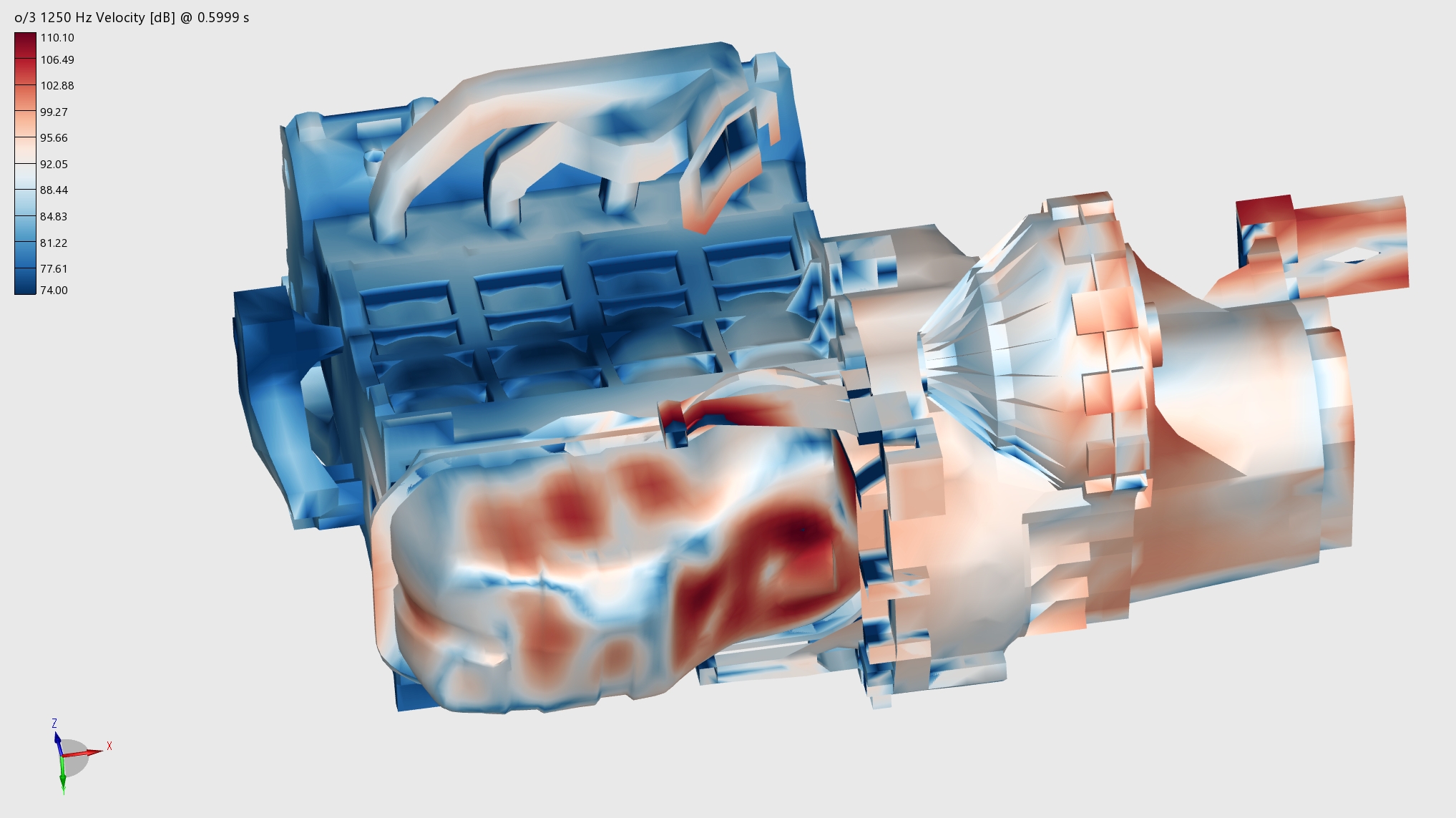}
    \caption{1250Hz \nicefrac{1}{3} octave}
    \label{fig:1250_linear}
\end{subfigure}
\caption{Velocity levels at the other three critical \nicefrac{1}{3} octaves.}
\label{fig:total_critical_third_octaves}
\end{figure*}

\subsection{Detailed analysis}
\par 
In order to  
{quantify} the limits violation, after qualitative exploration by means of visualization, we use \deleted{the} tooltips (Fig.~\ref{fig:velocity_levels_distributions}).
By ho\deleted{o}vering the pointer over the drill-down's top-left pane, we read just +0.1\,dB and +1.3\,dB over the acceptable limits for the 400\,Hz and the 1250\,Hz \nicefrac{1}{3} octaves, respectively, while we identify a quite significant excess of +5.4\,dB and +5.7\,dB in the 500\,Hz and the 630\,Hz \nicefrac{1}{3} octaves\deleted{, respectively}.  

\par   
Switching off the limits/show option, we depict the calculated integral velocity level values. 
The highest integral velocity levels occur below 200\,Hz: at the 67\,Hz \nicefrac{1}{3} octave, followed by the 80 and 125\,Hz \nicefrac{1}{3} octaves.
This can also be seen for the discrete velocity levels as depicted in Fig.~\ref{fig:cell_velocity_levels_wrt_colored}. 
These bands are outside the frequency range that is relevant for the noise analysis (the according limits are defined for the 215\,Hz\,\,--\,\,3150\,Hz range) as the human ear is not that sensitive at lower frequencies (the identified high velocity levels could become a subject of the vibration analysis, though).  
Accordingly, the analysis focus remained on the ``noise'' frequency bands in which the integral velocity levels are found to exceed the acceptable limits: 400, 500, 630 and 1250\,Hz.
%
%
\par 
The distribution of the 
{discrete} velocity levels (area-weighted) for a selection of frequency bands (octaves, \nicefrac{1}{3} octaves or harmonics) as well as of engine regions can be visualized using \replaced{boxplots}{box-plots} (Fig.~\ref{fig:boxplot}).  
The discrete velocity level distributions can be also shown against the discrete velocity level acceptance categories.
%
%
\par 
The matrix overview pane of the drill-down view (Fig.~\ref{fig:overview_discrete_limits}) can also show how much area of each engine region satisfies discrete velocity level limits for each frequency band.  
From above we know that the integral velocity levels for the whole engine (TOTAL region) are the highest at the 400\,Hz, 500\,Hz, 630\,Hz and 1250\,Hz \nicefrac{1}{3} octaves.  
From the pane we see that the main contributors to these high values at the specified frequency bands are the velocity levels at the LEFT, RIGHT, and BOTTOM regions.  
As the heights of the rectangles are proportional to region areas, we see that among the three regions the most critical one is the BOTTOM region.
\par 
This information can also be seen in detail in the \replaced{boxplots}{box-plots} (Fig.~\ref{fig:boxplot_split}) using the split option that depicts the discrete velocity levels distributions for a selection of the critical engine regions\,/\,frequency bands, from left to right: BOTTOM, LEFT, RIGHT.
\par 
In addition to the qualitative information in the overview matrix pane of the  drill-down view, we get also numerical values using the frequency band details
pane (Fig.~\ref{fig:geo3d_multi}) by selecting the critical regions\,/\,frequency bands in the overview matrix pane. 
We read that for the BOTTOM region in the critical frequency bands 50\,--\,70\,\% of its area has discrete velocity levels above the acceptance level (yellowish or reddish).
\par 
To identify engine surface parts related to the critical regions\,/\,frequency bands 
{spatially} we use the 3D view.
In addition to showing the discrete velocity level values (Fig.~\ref{fig:400_linear}), we can enable the limits flag to color the engine surface according to discrete acceptance criteria limits (Fig.~\ref{fig:400_limits}). 
Also, we can emphasize critical engine regions by showing the discrete velocity level values using a non-linear color distribution (Fig.~\ref{fig:400_non_linear}). 
\par 
Supported by AVL IMPRESS\texttrademark{} 3D and AVL IMPRESS\texttrademark/M, we can animate an exaggerated engine surface deformation for the targeted frequency bands.  
The surface is colored according to velocity levels in dB.  
This way, we can better focus on the identified acoustic hot spots and examine their deformation character.

\subsection{Findings}

Our exemplary analysis lead to the following findings: 
\begin{enumerate}
\itemsep=0pt
\item
High velocity levels in the BOTTOM region due to bending of the front, left, rear and bottom sides of the oil pan (Fig.~\ref{fig:400_linear}).
\item
High velocity levels in the LEFT and RIGHT regions due to lateral swinging of the intake and exhaust manifolds (Fig.~\ref{fig:500_linear}). 
\item
BOTTOM as the most critical engine region at the 630\,Hz \nicefrac{1}{3} octave with monopole noise sources at the side walls of the oil pan, having high radiation efficiency (Fig.~\ref{fig:630_linear}). 
\item
A vibration of the oil pan at the 1250\,Hz \nicefrac{1}{3} octave shows an excitation of its higher modes, with less radiation efficiency (Fig.~\ref{fig:1250_linear}). 
\item
The intake and exhaust manifolds at all critical frequencies behave as so-called dipole noise sources, meaning their swinging has also a lower radiation efficiency (Fig.~\ref{fig:500_linear}). 
\deleted{
High velocity levels, calculated at the intake and exhaust manifolds,  depend on boundary conditions applied to these parts during the engine modeling phase; in case the applied boundary conditions were ``too weak'', the highest achieved velocities would have lower values and would be shifted toward higher frequencies.} 
\end{enumerate}
As \replaced{a}{the} result, in the next design iteration, we \added{would} focus on a reduction of oil pan velocity levels at the 500 and 630\,Hz \nicefrac{1}{3} octaves by optimizing either the \added{geometric} design or \replaced{its}{the} excitation.

%
%
%
%
%
%
%
%
%
\section{Discussion}
\label{sec:discussion}
Designing an interactive visualization solution for domain experts poses several challenges.  Commonly, automotive engineers have decades of experience in simulation, using certain forms of visualization on a daily basis.  Usually, those amount to static images, which they use to make design decisions. 
In NVH studies, the data is large and multi-faceted, addressing spectral and spatial aspects, posing significant challenges for the analysis.  
The important semantic structures of both the frequency and the spatial domain imply additional challenges.  
%
\par 
Analyzing such data at multiple levels of detail is tedious with standard tools.  
Still, it is essential to flexibly relate frequencies and spatial locations for a successful analysis.
Our solution provides an integrated approach to jointly studying spatial and frequency aspects, from the overview level to a detailed analysis.  
This amounts to a major improvement over the state of the art.  
Many additional details, like custom color scales, synced 3D views, easy cells selections, ranking based brushing, etc., further improve the impact of the new approach.
\par 
The participatory design and development process, from the first exchange between simulation and visualization to the eventually designed visualization tool, was long and required numerous discussions, clarifications, many versions of design, including multiple different color scales, etc.~ 
Looking back,\deleted{ however,} we realize how important it was to repeatedly discuss visual analytics principles with the domain experts.  
\deleted{After all, they were very receptive to suggestions regarding the visualisation. 
Similarly, visualization researchers need to know about the priorities of the domain experts.}
%
%
\par 
\added{Due to the limited time the experts could spend for the evaluation and the fact that the data has to be pre-processed for the prototype, our solution was so far only used for the data presented in the use case in Section~\ref{sec:evaluation}.}
The overall feedback for the final prototype was very positive.  
\replaced{Earlier,}{Beforehand} the domain experts had to use several tools for the analysis. They had to identify a problematic point in the Campbell diagram, then load corresponding results in \replaced{a}{the} charting tool, look at the velocity charts per cell for a certain frequency, and finally highlight the cell in the 3D result viewer.
The \added{new} ease of accessing details and identifying spatial locations of the problematic cells is a very appreciated feature. Their feedback on our integrated solution was very positive since it enables them to quickly browse through the whole data \deleted{set} while keeping the context and not needing to switch applications.
We did not formally measure the speed-up, but engineers stated that the new approach is much faster. One consequence of the speed-up is that they can check much more details now. They can now very easily get data for a certain harmonic or a frequency band and hence check much more data which is not \replaced{located at}{in} problematic points. This helps in getting insight and a better comprehension of the data. Finally, the experts also stressed how the feature-rich interactive approach helps them in understanding the complexities of the analyzed system, and makes the identification of problematic spots easier in both domains.  
They stated that they cannot wait to have the new approach\added{, including the data pre-processing,} implemented in their production software\added{, so that more data sets could be explored}. They gladly demonstrated \replaced{the prototype}{it} to other colleges in their department \added{to gain support for the integration into the production software. Thanks to the positive feedback from these demonstrations the process of integration has started}.

The new approach was used repeatedly to confirm what the domain experts knew from experience already.  
Still, often the only way of communicating their findings is through standardized reports. 
To ensure that such reports are accepted by their customers and that they can be compared with other reports, standard color scales and charts must be used. 
Changing well-established domain conventions takes long time, if it happens at all.
\added{We hope that the integration in the production software will aid this process among NVH engineers.}
\par 
It is often desirable to semantically aggregate data and to use a hierarchical approach to exploration.
Frequency bands are a standard way of doing aggregation of harmonics in acoustics, but other domains often have similar concepts.
Some of our solutions also could be used in other domains with similar types of spectral data.  

Spectral imaging is often used in \replaced{medicine}{medical} and geology\deleted{ domains}.
In both domains\added{, the} spatial component is \added{also very} important.
If both spatial and frequency components are present, it is necessary to provide simultaneous access to and visualization of both components.  
Furthermore, some of our visual encoding solutions, such as splitting of heatmap cells to show overview and detail or using a \replaced{two-tone coloring}{horizon plot} inspired approach to depict\added{ing} values in heatmaps, can be used in many application domains. 
%
%

\section{Conclusion}
\label{sec:conclusion}
We present results of a collaboration between NVH domain experts and visualization researchers, \added{i.e.,} a novel visual analysis \replaced{solution}{tool} for noise simulation data.  
Our approach enables a quick drill-down from octave frequency bands to the individual harmonics \deleted{levels} without loosing the overview of the hierarchical data structure. 
The exploratory data analysis takes place simultaneously in the frequency and spatial components views. 
We see the potential that other professionals, who are concerned with multi-faceted data including frequency and spatial aspects, can draw valuable inspiration from our here presented solution.
\par 
Future directions of our research include the extension of the proposed approach to ensemble simulations and its evaluation in the context of different simulation models.  
The internal combustion engine was selected here since the domain experts are familiar with it and it seemed the most appropriate choice for a use case.   
Including other views, and interactively analyzing the data in a more abstract manner, and by means of automatic analysis methods, also, represent another important avenue for future research.  
We plan to explore novel use cases from different application domains with similar data.
\section*{Acknowledgements} 
VRVis is funded by BMK, BMDW, Styria, SFG, Tyrol and Vienna Business Agency in the scope of COMET \deleted{- Competence Centers for Excellent Technologies }(879730) which is managed by FFG.
Parts of this work have been done in the context of CEDAS, i.e., the Center of Data Science at the University of Bergen, Norway.
Parts of this work have been supported by Virginia Tech Institute for Creativity, Arts, and Technology.
\bibliographystyle{abbrv-doi}

\bibliography{IEEE_VIS_2022_NVH}

\begin{thebibliography}{10}

\bibitem{Boyce-2012-a}
M.~P. Boyce.
\newblock Rotor dynamics.
\newblock In M.~P. Boyce, ed., {\em Gas Turbine Engineering Handbook}, chap.~5,
  pp. 215--250. Butterworth-Heinemann, Waltham, MA 02451, fourth ed., 2012.

\bibitem{Campbell-1922-a}
W.~Campbell.
\newblock The protection of steam turbine disk wheels from axial vibration.
\newblock In {\em Proceedings of the Spring meeting of the American Society of
  Mechanical Engineers}, 1924.

\bibitem{Di-Marco-2019-a}
F.~Di~Marco, R.~D'Amico, and F.~Ronzio.
\newblock Electric motor encapsulation design for improved {NVH}: a {CAE}-based
  approach.
\newblock In {\em Proceedings of the INTER-NOISE and NOISE-CON Congress and
  Conference}, number~7, pp. 2464--2474, 2019.

\bibitem{Bonneau-2003-a}
H.~Doleisch, M.~Gasser, and H.~Hauser.
\newblock Interactive feature specification for focus+context visualization of
  complex simulation data.
\newblock In {\em Proceedings of the Eurographics / IEEE VGTC Symposium on
  Visualization}, pp. 239--248, 2003. doi: {{%
10\hspace{.1pt}\discretionary{.}{%
}{.}\hspace{.4pt}2312\discretionary{/}{%
}{/}VisSym\discretionary{/}{%
}{/}VisSym03\discretionary{/}{%
}{/}239\discretionary{%
}{-}{-}248}}


\bibitem{Dominguez-2019-a}
X.~Dominguez, P.~Arboleya, P.~Mantilla-Perez, I.~El-Sayed, N.~Gimenez, and
  M.~A. DiazMillan.
\newblock Visual analytics-based computational tool for electrical distribution
  systems of vehicles.
\newblock In {\em Proceedings of the 2019 IEEE Vehicle Power and Propulsion
  Conference, VPPC}, pp. 1--5, Oct. 2019. doi: {{%
10\hspace{.1pt}\discretionary{.}{%
}{.}\hspace{.4pt}1109\discretionary{/}{%
}{/}VPPC46532\hspace{.1pt}\discretionary{.}{%
}{.}\hspace{.4pt}2019\hspace{.1pt}\discretionary{.}{%
}{.}\hspace{.4pt}8952440}}


\bibitem{Eirich-2021-a}
J.~Eirich, J.~Bonart, D.~J\"ackle, M.~Sedlmair, U.~Schmid, K.~Fischbach,
  T.~Schreck, and J.~Bernard.
\newblock {IRVINE}: A design study on analyzing correlation patterns of
  electrical engines.
\newblock {\em IEEE Transactions on Visualization and Computer Graphics}, 2021.
  doi: {{%
10\hspace{.1pt}\discretionary{.}{%
}{.}\hspace{.4pt}1109\discretionary{/}{%
}{/}TVCG\hspace{.1pt}\discretionary{.}{%
}{.}\hspace{.4pt}2021\hspace{.1pt}\discretionary{.}{%
}{.}\hspace{.4pt}3114797}}


\bibitem{Eirich-2020-a}
J.~Eirich, D.~J{\"{a}}ckle, T.~Schreck, J.~Bonart, O.~Posegga, and
  K.~Fischbach.
\newblock {VIMA}: Modeling and visualization of high dimensional machine sensor
  data leveraging multiple sources of domain knowledge.
\newblock In {\em Proceedings of the 2020 Visualization in Data Science, VDS},
  pp. 22--31, Oct. 2020. doi: {{%
10\hspace{.1pt}\discretionary{.}{%
}{.}\hspace{.4pt}1109\discretionary{/}{%
}{/}VDS51726\hspace{.1pt}\discretionary{.}{%
}{.}\hspace{.4pt}2020\hspace{.1pt}\discretionary{.}{%
}{.}\hspace{.4pt}00007}}


\bibitem{Hampl-2010-a}
N.~Hampl.
\newblock Advanced simulation techniques in vehicle noise and vibration
  refinement.
\newblock In X.~Wang, ed., {\em Vehicle Noise and Vibration Refinement},
  chap.~8, pp. 174--188. Woodhead Publishing, 2010. doi: {{%
10\hspace{.1pt}\discretionary{.}{%
}{.}\hspace{.4pt}1533\discretionary{/}{%
}{/}9781845698041\hspace{.1pt}\discretionary{.}{%
}{.}\hspace{.4pt}2\hspace{.1pt}\discretionary{.}{%
}{.}\hspace{.4pt}174}}


\bibitem{Harrison-2004-a}
M.~Harrison.
\newblock {\em Vehicle Refinement: Controlling Noise and Vibration in Road
  Vehicles}.
\newblock Elsevier Butterworth-Heinemann, Burlington, MA, 2004.

\bibitem{KH2012}
J.~Kehrer and H.~Hauser.
\newblock Visualization and visual analysis of multifaceted scientific data: A
  survey.
\newblock {\em IEEE Transactions on Visualization and Computer Graphics},
  19(3):495--513, 2013. doi: {{%
10\hspace{.1pt}\discretionary{.}{%
}{.}\hspace{.4pt}1109\discretionary{/}{%
}{/}TVCG\hspace{.1pt}\discretionary{.}{%
}{.}\hspace{.4pt}2012\hspace{.1pt}\discretionary{.}{%
}{.}\hspace{.4pt}110}}


\bibitem{Keim2008}
D.~Keim, G.~Andrienko, J.-D. Fekete, C.~G{\"o}rg, J.~Kohlhammer, and
  G.~Melan{\c{c}}on.
\newblock {\em Visual Analytics: Definition, Process, and Challenges}, pp.
  154--175.
\newblock Springer, 2008. doi: {{%
10\hspace{.1pt}\discretionary{.}{%
}{.}\hspace{.4pt}1007\discretionary{/}{%
}{/}978\discretionary{%
}{-}{-}3\discretionary{%
}{-}{-}540\discretionary{%
}{-}{-}70956\discretionary{%
}{-}{-}5\_7}}


\bibitem{Kniss-2001-a}
J.~Kniss, G.~Kindlmann, and C.~Hansen.
\newblock Interactive volume rendering using multi-dimensional transfer
  functions and direct manipulation widgets.
\newblock In {\em Proceedings of the 2001 IEEE Visualization Conference, IEEE
  VIS'01}, pp. 255--262. IEEE, 21--26~Oct. 2001.

\bibitem{Langer-2021-a}
T.~Langer and T.~Meisen.
\newblock Visual analytics for industrial sensor data analysis.
\newblock In {\em Proceedings of the 23rd International Conference on
  Enterprise Information Systems, ICEIS}, pp. 584--593, 2021. doi: {{%
10\hspace{.1pt}\discretionary{.}{%
}{.}\hspace{.4pt}5220\discretionary{/}{%
}{/}0010399705840593}}


\bibitem{Mandke-2019-a}
D.~Mandke, D.~Ghaisas, S.~Pawar, and S.~Suh.
\newblock Numerical prediction and verification of noise radiation
  characteristics of diesel engine block.
\newblock In {\em Proceedings of the \ Noise and Vibration Conference {\&}
  Exhibition}. SAE International, 2019. doi: {{%
10\hspace{.1pt}\discretionary{.}{%
}{.}\hspace{.4pt}4271\discretionary{/}{%
}{/}2019\discretionary{%
}{-}{-}01\discretionary{%
}{-}{-}1591}}


\bibitem{Neeman-2005-a}
A.~Neeman, B.~Jeremic, and A.~Pang.
\newblock Visualizing tensor fields in geomechanics.
\newblock In {\em Proceedings of the 2005 IEEE Visualization Conference, IEEE
  VIS'05}, pp. 35--42. IEEE, 23--28~Oct. 2005.

\bibitem{Prajith-2020-a}
J.~Prajith and V.~Sagade.
\newblock {NVH} full vehicle development~-- virtual simulation process for low
  frequency structure-borne regions.
\newblock In {\em Proceedings of the WCX SAE World Congress Experience}. SAE
  International, 2020. doi: {{%
10\hspace{.1pt}\discretionary{.}{%
}{.}\hspace{.4pt}4271\discretionary{/}{%
}{/}2020\discretionary{%
}{-}{-}01\discretionary{%
}{-}{-}1266}}


\bibitem{CMV_STAR}
J.~C. Roberts.
\newblock State of the art: Coordinated \& multiple views in exploratory
  visualization.
\newblock In {\em Proceedings of the Fifth International Conference on
  Coordinated and Multiple Views in Exploratory Visualization, CMV 2007}, pp.
  61--71, 2007. doi: {{%
10\hspace{.1pt}\discretionary{.}{%
}{.}\hspace{.4pt}1109\discretionary{/}{%
}{/}CMV\hspace{.1pt}\discretionary{.}{%
}{.}\hspace{.4pt}2007\hspace{.1pt}\discretionary{.}{%
}{.}\hspace{.4pt}20}}


\bibitem{Saito-2005-a}
T.~Saito, H.~Miyamura, M.~Yamamoto, H.~Saito, Y.~Hoshiya, and T.~Kaseda.
\newblock Two-tone pseudo coloring: Compact visualization for one-dimensional
  data.
\newblock In {\em Proceedings of the 2005 IEEE Symposium on Information
  Visualization, INFOVIS 2005}, pp. 173--180. IEEE, 23--25~Oct. 2005.

\bibitem{TwoTone}
T.~Saito, H.~N. Miyamura, M.~Yamamoto, H.~Saito, Y.~Hoshiya, and T.~Kaseda.
\newblock Two-tone pseudo coloring: compact visualization for one-dimensional
  data.
\newblock In {\em Proceedings of the IEEE Symposium on Information
  Visualization, INFOVIS}, pp. 173--180, 2005. doi: {{%
10\hspace{.1pt}\discretionary{.}{%
}{.}\hspace{.4pt}1109\discretionary{/}{%
}{/}INFVIS\hspace{.1pt}\discretionary{.}{%
}{.}\hspace{.4pt}2005\hspace{.1pt}\discretionary{.}{%
}{.}\hspace{.4pt}1532144}}


\bibitem{Sedlmair-2011-a}
M.~Sedlmair, P.~Isenberg, D.~Baur, M.~Mauerer, C.~Pigorsch, and A.~Butz.
\newblock Cardiogram: Visual analytics for automotive engineers.
\newblock In {\em Proceedings of the SIGCHI Conference on Human Factors in
  Computing Systems}, pp. 1727--1736. Association for Computing Machinery,
  2011. doi: {{%
10\hspace{.1pt}\discretionary{.}{%
}{.}\hspace{.4pt}1145\discretionary{/}{%
}{/}1978942\hspace{.1pt}\discretionary{.}{%
}{.}\hspace{.4pt}1979194}}


\bibitem{Shaik-Mohammad-2017-a}
A.~B. {Shaik Mohammad}, R.~Vijayakumar, and N.~R. Panduranga.
\newblock Structure borne noise optimization of diesel engine by simulation.
\newblock In {\em Proceedings of the International Conference on Advances in
  Design, Materials, Manufacturing and Surface Engineering for Mobility}. SAE
  International, 2017. doi: {{%
10\hspace{.1pt}\discretionary{.}{%
}{.}\hspace{.4pt}4271\discretionary{/}{%
}{/}2017\discretionary{%
}{-}{-}28\discretionary{%
}{-}{-}1944}}


\bibitem{Shaik-Mohammad-2019-a}
A.~B. {Shaik Mohammad}, R.~Vijayakumar, and N.~R. Panduranga.
\newblock Vibro-acoustic optimization of 3 cylinder diesel engine components
  for lower sound radiation using finite element techniques.
\newblock In {\em Proceedings of the Symposium on International Automotive
  Technology}. SAE International, 2019. doi: {{%
10\hspace{.1pt}\discretionary{.}{%
}{.}\hspace{.4pt}4271\discretionary{/}{%
}{/}2019\discretionary{%
}{-}{-}26\discretionary{%
}{-}{-}0189}}


\bibitem{Shneiderman-1996-a}
B.~Shneiderman.
\newblock The eyes have it: A task by data type taxonomy for information
  visualizations.
\newblock In {\em Proceedings 1996 IEEE Symposium on Visual Languages}, pp.
  336--343. IEEE, 3--6~Sept. 1996.

\bibitem{Souksavanh-2020-a}
V.~Souksavanh and Y.~Liu.
\newblock {NVH} data analytics and its application in vehicle rating.
\newblock In {\em Proceedings of the 2020 IEEE 7th International Conference on
  Industrial Engineering and Applications, ICIEA}, pp. 287--292. IEEE, 2020.
  doi: {{%
10\hspace{.1pt}\discretionary{.}{%
}{.}\hspace{.4pt}1109\discretionary{/}{%
}{/}ICIEA49774\hspace{.1pt}\discretionary{.}{%
}{.}\hspace{.4pt}2020\hspace{.1pt}\discretionary{.}{%
}{.}\hspace{.4pt}9101968}}


\bibitem{Stevens-2007-a}
J.~A. Stevens.
\newblock Visualization of complex automotive data: A tutorial.
\newblock {\em IEEE Computer Graphics and Applications}, 27(6):80--86, 2007.
  doi: {{%
10\hspace{.1pt}\discretionary{.}{%
}{.}\hspace{.4pt}1109\discretionary{/}{%
}{/}MCG\hspace{.1pt}\discretionary{.}{%
}{.}\hspace{.4pt}2007\hspace{.1pt}\discretionary{.}{%
}{.}\hspace{.4pt}161}}


\bibitem{tukey1977exploratory}
J.~W. Tukey et~al.
\newblock {\em Exploratory data analysis}, vol.~2.
\newblock Reading, MA, 1977.

\bibitem{Vernon-Bido-2015-a}
D.~Vernon-Bido, A.~Collins, and J.~Sokolowski.
\newblock Effective visualization in modeling {\&} simulation.
\newblock In {\em Proceedings of the 48th Annual Simulation Symposium}, pp.
  33--40. Society for Computer Simulation International, San Diego, 2015.

\bibitem{Wang-2010-a}
X.~Wang.
\newblock Rationale and history of vehicle noise and vibration refinement.
\newblock In X.~Wang, ed., {\em Vehicle Noise and Vibration Refinement},
  chap.~1, pp. 3--17. Woodhead Publishing, 2010. doi: {{%
10\hspace{.1pt}\discretionary{.}{%
}{.}\hspace{.4pt}1533\discretionary{/}{%
}{/}9781845698041\hspace{.1pt}\discretionary{.}{%
}{.}\hspace{.4pt}1\hspace{.1pt}\discretionary{.}{%
}{.}\hspace{.4pt}3}}


\bibitem{Zheng-2002-a}
X.~Zheng and A.~Pang.
\newblock Volume deformation for tensor visualization.
\newblock In {\em Proceedings of the 2002 IEEE Visualization Conference, IEEE
  VIS'02}, pp. 379--386. IEEE, 27~Oct.--1~Nov. 2002.

\end{thebibliography}
\end{document}